\global\def\draftcontrol{0}

   \def\versionno{ open hypersurfaces }
\catcode`\@=11

\expandafter\ifx\csname draftcontrol\endcsname\relax\global\def\draftcontrol{0} 
\fi 

{\count255=\time\divide\count255 by 60 
\xdef\hourmin{\number\count255} 
\multiply\count255 by-60\advance\count255 by\time 
\xdef\hourmin{\hourmin:\ifnum\count255<10 0\fi\the\count255}} 
\def\draftdate{\number\month/\number\day/\number\year\ \ \ \hourmin } 

\newcommand\makepapertitle{\par

  \begingroup 
    \renewcommand\thefootnote{\@fnsymbol\c@footnote}% 
    \def\@makefnmark{\rlap{\@textsuperscript{\normalfont\@thefnmark}}}% 
    \long\def\@makefntext##1{\parindent 1em\noindent 
            \hb@xt@1.8em{% 
                \hss\@textsuperscript{\normalfont\@thefnmark}}##1}% 
     \newpage 
     \global\@topnum\z@   % Prevents figures from going at top of page. 
     \@makepapertitle 
     \thispagestyle{empty}\@thanks 
  \endgroup 
  \setcounter{footnote}{0}% 
  \global\let\thanks\relax 
  \global\let\makepapertitle\relax 
  \global\let\@makepapertitle\relax 
  \global\let\@thanks\@empty 
  \global\let\@author\@empty 
  \global\let\@date\@empty 
  \global\let\@title\@empty 
  \global\let\title\relax 
  \global\let\author\relax 
  \global\let\date\relax 
  \global\let\and\relax 
  \def\version{\let\version\@version\@gobble} 
} 
\def\@makepapertitle{% 
  \newpage 
   \ifnum\draftcontrol=1 {} 
   \version\versionno 
   \vskip 5.5em% 
   \else 
   \hfill\hbox to 3.5cm {\parbox{4.5cm}{\@pubnum}\hss}% 
   \vskip 6.5em% 
   \fi 
   \begin{center}% 
   \let \footnote \thanks 
      {\hskip -0\textwidth \hbox to 1\textwidth% 
        {\centerline{\Large\bf{\noindent\@title}}}}% 
     \vskip 2em% 
     {\normalsize%\large 
       \lineskip .5em% 
       \begin{tabular}[t]{c}% 
         \@author 
       \end{tabular}\par}% 
     \vskip 1.5em% 
     {\@bstract}% 
     \end{center}% 
     \vfill
     \@date%
     \vskip 1.5em%
   \par 
} 

\gdef\@pubnum{} 
\def\pubnum#1{% 
  \gdef\@pubnum{#1}} 

\gdef\@bstract{} 
\def\Abstract#1{% 
  \gdef\@bstract{% 
   \parbox{\textwidth-0pc}{% 
   \centerline{\bf Abstract}\penalty1000 
   \noindent%\abstractfont \baselineskip=12pt 
   \renewcommand\baselinestretch{1.0} 
   {#1}}} 
} 

\gdef\@email{}
\def\email#1{%
   \gdef\@email{%
   Email: {\tt #1}}
}

\def\ps@paper{\let\@mkboth\@gobbletwo% 
     \ifnum\draftcontrol=1 
        \def\@oddfoot{\hbox to \textwidth{\tiny \versionno \hfil\tiny\draftdate}% 
        \hskip -\textwidth \hbox to \textwidth{\hfil\rm\thepage\hfil}}% 
     \else\def\@oddfoot{\hbox to \textwidth{\hfil\rm\thepage\hfil}} 
     \fi 
     \let\@evenfoot\@oddfoot 
} 
\def\body{\clearpage 
          \pagestyle{paper} 
        } 
\newenvironment{acknowledgments}{% 
\vskip 3.25ex 
\addcontentsline{toc}{section}{Acknowledgments}
\noindent {\bf Acknowledgments} 
} 

\def\@version#1{\ifnum\draftcontrol=1 
\typeout{}\typeout{#1}\typeout{} 
\vskip3mm\centerline{\hbox{\fbox{\normalsize{\tt DRAFT -- #1 -- } 
                   {\draftdate}}}}\vskip3mm 
\fi} 
\let\version\@version 
\long\def\eqlabel#1{\ifnum\draftcontrol=1 
                    \tag@false  % there are some problems with multline without this 
                    \tag*{(\theequation) \hbox to -0.2cm{\hspace{0cm}\small{#1}\hss}} 
                    \refstepcounter{equation}  
                    \edef\@currentlabel{\theequation} 
                    \ltx@label{#1}          % use old LaTeX \label instead of new definition 
                    \else 
                    \label{#1} 
                    \fi 
                    } 
\let\st@bibitem\@bibitem 
\let\st@lbibitem\@lbibitem 
\ifnum\draftcontrol=1 
  \def\@bibitem#1{% 
    \st@bibitem{#1}\a@@label{#1}\ignorespaces} 
  \def\@lbibitem[#1]#2{% 
    \st@lbibitem[#1]{#2}\a@@label{#2}\ignorespaces} 
  \def\a@@label#1{% 
    \gdef\a@lab{\smash{\normalfont\small#1}} 
    \ifvmode 
      \if@inlabel 
        \global\setbox\@labels\hbox{% 
          \llap{\a@lab\let\a@lab\relax 
                \kern\@totalleftmargin\kern\marginparsep}% 
          \box\@labels}% 
      \fi 
    \fi} 
\fi 
\documentclass[12pt,letterpaper]{article} 
\usepackage{amsmath,bm,amsfonts,amssymb,array,calc,amsthm,rotating}
\usepackage{epsfig,psfrag} 
\usepackage{graphicx}
\usepackage{color}
\usepackage[colorlinks=false]{hyperref}
\renewcommand\baselinestretch{1.25} 
\renewcommand\section{\@startsection {section}{1}{\z@}% 
                                   {-3.5ex \@plus -1ex \@minus -.2ex}% 
                                   {2.3ex \@plus.2ex}% 
                                   {\normalfont\large\bfseries}} 
\renewcommand\subsection{\@startsection{subsection}{2}{\z@}% 
                                   {-3.25ex\@plus -1ex \@minus -.2ex}% 
                                   {1.5ex \@plus .2ex}% 
                                   {\normalfont\normalsize\bfseries}} 
\renewcommand\subsubsection{\@startsection{subsubsection}{3}{\z@}% 
                                   {-3.25ex\@plus -1ex \@minus -.2ex}% 
                                   {1.5ex \@plus .2ex}% 
                                   {\normalfont\normalsize\it}} 
\renewcommand\paragraph{\@startsection{paragraph}{4}{\z@}% 
                                   {-3.25ex\@plus -1ex \@minus -.2ex}% 
                                   {1.5ex \@plus .2ex}% 
                                   {\normalfont\normalsize\bf}} 
\renewcommand\subparagraph{\@startsection{subparagraph}{5}{\z@}% 
                                   {-1.25ex\@plus -1ex \@minus -.2ex}% 
                                   {0ex \@plus .2ex}% 
                                   {\normalfont\normalsize\it}}

\numberwithin{equation}{section}

\long\def\@makecaption#1#2{%
  \vskip\abovecaptionskip
  \sbox\@tempboxa{{\bf #1:} #2}%
  \ifdim \wd\@tempboxa >\hsize
    {\small\bf #1:} {\small #2}\par
  \else
    \global \@minipagefalse
    \hb@xt@\hsize{\hfil\box\@tempboxa\hfil}%
  \fi
  \vskip\belowcaptionskip}
\renewcommand*\l@section[2]{%
  \ifnum \c@tocdepth >\z@
    \addpenalty\@secpenalty
    \addvspace{.5em \@plus\p@}%
    \setlength\@tempdima{1.5em}%
    \begingroup
      \parindent \z@ \rightskip \@pnumwidth
      \parfillskip -\@pnumwidth
      \leavevmode \bfseries
      \advance\leftskip\@tempdima
      \hskip -\leftskip
      #1\nobreak\hfil \nobreak\hb@xt@\@pnumwidth{\hss #2}\par
    \endgroup
  \fi}
\renewcommand*\l@subsection{\addvspace{.0em \@plus\p@}\@dottedtocline{2}{1.5em}{2.3em}}
\renewcommand*\l@subsubsection{\addvspace{-.2em \@plus\p@}\@dottedtocline{3}{3.8em}{3.2em}}

\def\hepth#1{\href{http://xxx.arxiv.org/abs/hep-th/#1}{{arXiv:hep-th/#1}}}

\def\math#1{\href{http://xxx.arxiv.org/abs/math/#1}{{arXiv:math/#1}}}

\def\mathsg#1{\href{http://xxx.arxiv.org/abs/math.SG/#1}{{arXiv:math.sg/#1}}}
\def\mathag#1{\href{http://xxx.arxiv.org/abs/math.AG/#1}{{arXiv:math.ag/#1}}}

\def\arxiv#1#2{\href{http://xxx.arxiv.org/abs/#1}{{arXiv:#1 [#2]}}}

\definecolor{refcol}{rgb}{0.2,0.2,0.8}
\definecolor{eqcol}{rgb}{.6,0,0}
\definecolor{purple}{cmyk}{0,1,0,0}

\gdef\@citecolor{refcol}
\gdef\@linkcolor{eqcol}
\def\colorlinkspurple{\gdef\@urlcolor{purple}}
\def\colorlinksblue{\gdef\@urlcolor{blue}}
\def\colorlinksred{\gdef\@urlcolor{red}}

\def\ie{{\it i.e.}} 
\def\eg{{\it e.g.}} 

\def\cf{{\it cf.}}

\def\revise#1       {\raisebox{-0em}{\rule{3pt}{1em}}% 
                     \marginpar{\raisebox{.5em}{\vrule width3pt\ 
                     \vrule width0pt height 0pt depth0.5em 
                     \hbox to 0cm{\hspace{0cm}{% 
                     \parbox[t]{4em}{\raggedright\footnotesize{#1}}}\hss}}}}

\def\calf         {{\cal F}}

\def\caln         {{\cal N}} 
\def\calo         {{\cal O}}

\def\calt         {{\cal T}}

\def\calw         {{\cal W}}

\def\projective   {{\mathbb P}} 
 
\def\reals        {{\mathbb R}} 
\def\zet          {{\mathbb Z}} 

\def\RP{\reals\projective}

\def\del          {\partial} 
 
\def\ee           {{\it e}} 
\def\ii           {{\it i}}

\def\Im           {{\rm Im\hskip0.1em}}

\def\sqr#1#2{{\vcenter{\vbox{\hrule height.#2pt   
 \hbox{\vrule width.#2pt height#1pt \kern#1pt 
 \vrule width.#2pt}\hrule height.#2pt}}}}

\renewcommand{\t}{\tilde}
\renewcommand{\c}{\bar}
\newcommand{\h}{\hat}

\newcommand{\C}{\mathbb C}
\newcommand{\Z}{\mathbb Z}

\renewcommand{\L}{\mathcal L}
\renewcommand{\P}{\mathbb P}
\newcommand{\R}{\mathbb R}

\newcommand{\T}{\mathcal T}

\newcommand{\PF}{\mathcal L}

\newcommand{\G}{\Gamma}
\newcommand{\p}{\psi}
\newcommand{\w}{\omega}

\newcommand{\e}{\epsilon}
\newcommand{\vp}{\varpi}
\renewcommand{\a}{\alpha}
\renewcommand{\th}{\theta}

\renewcommand{\w}{\wedge}
\newcommand{\D}{\partial}
\newcommand{\beq}{\begin{equation}}
\newcommand{\eq}{\end{equation}}

\def\ker{\mathop{\rm Ker}}
\def\Ord{\mathop{\rm Ord}}

\catcode`\@=12 

\begin{document} 

%%% 
%%%%%% text starts here 
%%%%%%%%% 

\title{Real Mirror Symmetry for One-Parameter Hypersurfaces}

\pubnum{
arXiv:0805.0792\\
CERN-PH-TH/2008-091\\
LMU-ASC 24/08
}

\date{May 2008}

\author{
Daniel Krefl$^{a}$ and Johannes Walcher$^{b,}$\footnote{On leave from:
Institute for Advanced Study, Princeton, NJ, USA} \\[0.2cm]
\it ${}^{a}$ Arnold Sommerfeld Center for Theoretical Physics, LMU Munich, Germany \\
\it PH-TH Division, CERN, Geneva, Switzerland \\[0.2cm]
\it $^{b}$ Institute for Theoretical Physics, ETH Zurich, Switzerland 
}

\Abstract{
We study open string mirror symmetry for one-parameter Calabi-Yau hypersurfaces in
weighted projective space. We identify mirror pairs of D-brane configurations, derive 
the corresponding inhomogeneous Picard-Fuchs equations, and solve for the domainwall
tensions as analytic functions over moduli space. Our calculations exemplify several
features that had not been seen in previous work on the quintic or local Calabi-Yau 
manifolds. We comment on the calculation of loop amplitudes.}

\makepapertitle

\body

\version\versionno

\vskip 1em

\tableofcontents
%\newpage

\section{Introduction and Overview}

The purpose of this paper is to extend recent progress in open string mirror symmetry
for the quintic \cite{Walcher:2006rs,Morrison:2007bm} to the class of Calabi-Yau 
hypersurfaces, $X$, in weighted projective space, with one-dimensional K\"ahler 
moduli space, \ie, $h_{11}(X)=1=h_{21}(Y)$, where $Y$ is the mirror manifold of $X$.

There are three models of this type (excluding the quintic), characterized by the five
positive integer weights $(\nu_1,\ldots,\nu_5)$, such that $k:=\sum \nu_i$ is divisible 
by each of the $\nu_i$ (Gepner models), and all five mutually coprime. We will denote 
$k/\nu_i=:h_i$. These models were considered in the early days of mirror symmetry 
\cite{Morrison:1991cd,Klemm:1992tx,Font:1992uk} as the simplest class of examples to 
which to extend the original computation of Candelas et al.\ \cite{cdgp} on the quintic.

The manifolds of the A-model, $X^{(k)}$, are hypersurfaces of degree $k$ in weighted 
projective space $\P^4({\nu_1,\ldots,\nu_5})$:
\beq\eqlabel{modelseq0}
\begin{split}
X^{(6)} &\subset \P^4(1,1,1,1,2)~, \\
X^{(8)} &\subset \P^4(1,1,1,1,4)~, \\
X^{(10)} &\subset \P^4(1,1,1,5,2)~. \\
\end{split}
\end{equation}
The corresponding mirror manifolds, $Y^{(k)}$, are resolutions of quotients of specific 
one-parameter families of degree $k$ hypersurfaces by the group $G=\hat G/\zet_k$, where
$\hat G=\ker\bigl(\prod_i \zet_{h_i}\to \zet_k\bigr)$ is the Greene-Plesser orbifold group.
\beq\eqlabel{modelseq1}
\begin{split}
Y^{(6)}&: 
\frac{1}{6}\left(x^6_1+x^6_2+x^6_3+x^6_4+2x^3_5\right)-\psi x_1x_2x_3x_4x_5=0~,\\
Y^{(8)}&:
\frac{1}{8}\left(x^8_1+x^8_2+x^8_3+x^8_4+4x^2_5\right)-\psi x_1x_2x_3x_4x_5=0~,\\
Y^{(10)}&: 
\frac{1}{10}\left(x^{10}_1+x^{10}_2+x^{10}_3+2x^5_4+5x^2_5\right)-\psi x_1x_2x_3x_4x_5=0~.
\end{split}
\eq

For compact Calabi-Yau manifolds, the only systematic construction of D-branes of the A-model
(Lagrangian submanifolds) is as the fixed point set of an anti-holomorphic involution for some
choice of complex structure on $X^{(k)}$. The Fermat point $\sum x_i^{h_i}=0$ in complex structure 
moduli space is the most convenient for comparison with boundary conformal field theory 
and derivation of the mirror configurations (see below). For $X^{(6)}$ and $X^{(10)}$,
the Lagrangians and their worldvolume theories are qualitatively very similar to those on
the quintic, see section \ref{configs}. For $X^{(8)}$, we find that for certain choices of 
anti-holomorphic involution, the fixed point set consists of {\it disconnected components} 
in the {\it same homology class}. As a consequence, the worldvolume theory has vacua corresponding 
to wrapping on different Lagrangian submanifolds that cannot be continuously connected through
Lagrangian families. In this case, the tension of BPS domainwalls between the vacua receives 
corrections from both {\it open and closed} string worldsheet instantons. Only when all 
effects are combined do we obtain a physically sensible domainwall spectrum.
Although this phenomenon is not unexpected in general, it had not been 
seen previously in explicit examples on either the quintic or local Calabi-Yau manifolds.

On the mirror side, the most convenient (and complete) description of B-type D-branes on $Y^{(k)}$ 
is as graded, $\hat G$-equivariant matrix factorizations of the hypersurface polynomial, $W^{(k)}$, 
viewed as Landau-Ginzburg superpotential \cite{stability}. The basic algorithm for working out the 
configurations mirror to the real slices of $X^{(k)}$ is described in \cite{bhhw}. We will follow 
this procedure and match the vacuum structure with that seen in the A-model.

For certain other choices of anti-holomorphic involution, also on $X^{(8)}$, the Lagrangian 
submanifold has a non-zero first Betti number, and hence a classical deformation space. It is 
of interest to ask whether this moduli space is lifted by quantum effects (worldsheet instantons). 
We cannot at the moment answer this question from A-model considerations. However, if our mirror 
proposal is correct, the B-model results indicate that this moduli space in fact persists at the 
quantum level, \ie, no superpotential is generated for the corresponding chiral field. 

We then turn in section \ref{picardfuchs} to the computation of more refined invariant information, 
namely the tension of BPS domainwalls, or superpotential differences, between the various brane vacua. 
As explained in \cite{Morrison:2007bm}, the appropriate mathematical concept is that of a Hodge
theoretic {\it normal function}. In the B-model, it can be represented geometrically as an integral 
of the holomorphic three-form over a three-chain suspended between homologically equivalent 
holomorphic curves. The curves representing the brane vacua of our interest can be determined 
algorithmically from the matrix factorization via the algebraic second Chern class. The chain 
integral then satisfies an inhomogeneous version of the Picard-Fuchs differential equation 
governing closed string mirror symmetry, with an inhomogeneous term that can be computed 
explicitly from the curve and the Griffiths-Dwork algorithm. This Abel-Jacobi type method 
developped in \cite{Morrison:2007bm} is similar in spirit to the computations in local 
geometries \cite{agva,akv,lmw}.

With the inhomogeneous Picard-Fuchs equation in hand, we can then compute the fully quantum 
corrected domainwall tension over the entire closed string moduli space, see section 
\ref{monodromy}. We will check integrality of all requisite monodromy matrices, as 
well as the spectrum of tensionless domainwalls, expected from the matrix factorization 
considerations. By expanding around large complex structure/large volume, we obtain numerical 
predictions for the number of disks ending on the Lagrangians of the A-model, consistent with 
Ooguri-Vafa integrality \cite{oova}. 

In the final section \ref{conclude}, we will (before concluding) briefly discuss the computation 
of loop amplitudes in topological string backgrounds that include the D-branes we have studied. 
These computations, the details of which we leave for the future, use the extension 
\cite{openbcov,unorbcov} of the BCOV holomorphic anomaly equations \cite{bcov1,bcov2}, and 
yield further numerical predictions as well as additional consistency checks. Again, the 
most interesting example is $X^{(8)}$, where the disconnected fixed point set (as orientifold 
plane) allows for a variety of tadpole cancelling D-brane configurations.

\section{D-brane Configurations}
\label{configs}

The Fermat polynomial defining the A-model
\begin{equation}
\eqlabel{defining}
\sum_{i=1}^5 x_i^{h_i} = 0~, 
\end{equation}
where $h_i:=k/\nu_i$, is invariant under anti-holomorphic involutions acting as
\beq
\eqlabel{modelseq2}
x_i\rightarrow \phi_i^{M_i}\c x_i~,
\eq
where $\phi_i=e^{\frac{2\pi \ii \nu_i}{k}}$ are phases with $\nu_i$ the weights of the ambient 
weighted $\P^4\supset X^{(k)}$. The $M_i$ are integer, but the sets $(M_i)$ and
$(M_i+\nu_i)$ define the same involution by projective identification. The fixed-point 
loci, $L_{[M]}^{(k)}$, of the involutions \eqref{modelseq2} are special Lagrangian 
submanifolds of $X^{(k)}$, and can be parameterized explicitly by $x_i=\phi_i^{M_i/2}y_i$, 
with $y_i$ real. When $h_i$ is odd, two involutions differing only in $M_i$ are equivalent 
(though not identical) under the global symmetry group $\zet_{h_i}$, hence the corresponding 
$L_{[M]}^{(k)}$ are isomorphic. When $h_i$ is even, we have to distinguish whether $M_i$ is 
even or odd, which yields a sign in the equation determining the real locus,
\begin{equation}
\eqlabel{reallocus}
\sum_{i=1}^5 (-1)^{M_i} y_i^{h_i} = 0~.
\end{equation}
Again, for $h_i$ odd, $M_i$ is equivalent to $M_i+1$ by changing $y_i\to -y_i$.

\subsection{\texorpdfstring{$X^{(6)}$ and $X^{(10)}$ in the A-model}{X(6) and X(10) in the A-model}}

When at least one $h_i$ is odd, say $h_5$, we can solve \eqref{reallocus} uniquely over the 
reals for $y_5$, and identify
\begin{equation}
L^{(k)}_{[M]}\cong \{ (y_1,\ldots,y_4)\neq (0,0,0,0)\}/\reals^* \cong \RP^3~.
\end{equation}
Thus, 
\beq
L_{[M]}^{(6)}\cong\R\P^3, \qquad L_{[M]}^{(10)}\cong\R\P^3~.
\eq 
for all $M$. The vacuum structure of a D-brane wrapped on $L^{(6)}_{[M]}$ or $L^{(10)}_{[M]}$ 
(think of a D6 or D4-brane in type IIA) is therefore very similar to the quintic 
\cite{Walcher:2006rs}. In detail, since $H_1(\R\P^3,\Z)=\Z_2$, there is a discrete 
choice of Wilson line on the D-brane wrapping the $\R\P^3$ such that the worldvolume 
gauge theory will have two vacua, which we will parameterize by the discrete modulus 
$\sigma=\pm1$. A BPS domainwall separating the two vacua can be obtained by wrapping a 
(D4 or D2-) brane on a holomorphic disk in $X^{(k)}$ with boundary on the non-trivial 
one-cycle in $\R\P^3$ and with the remaining dimensions located in 
space-time \cite{Walcher:2006rs}. The corresponding domainwall tension, which we will denote
by $\T_A$, is the basic holomorphic observable associated with the D-brane configuration. 
At large volume, $\T_A$ clearly scales as $\T_A\sim t$, where $t$ is the K\"ahler modulus.
There are then quantum corrections to $\T_A$ due to worldsheet (disk) instantons.
Monodromy considerations around $\Im(t)\to +\infty$ identical to those on the quintic 
(which we will review momentarily) lead us to expect an expansion\footnote{The basic 
fact is the exact sequence $H_2(X;\zet)\to H_2(X,L;\zet)\to H_1(L;\zet)$ in which
the generator of $H_2(X,L;\zet)\cong\zet$ is mapped to the non-trivial class in 
$H_1(L;\zet)\cong H^2(L;\zet)\cong\zet_2$.}
\begin{equation}
\eqlabel{TA610}
\T_A = \frac{t}{2}+\Bigl(\frac 14 +\frac{1}{2\pi^2} \sum_{d\;{\rm odd}}\tilde n_d q^{d/2}\Bigr)~,
\end{equation}
where $q\equiv \exp(2\pi\ii t)$ and $\tilde n_d$ are the open Gromov-Witten invariants
counting holomorphic maps from the disk to $X^{(k)}$ with boundary on $L^{(k)}_{[M]}$.

The reasoning that leads to the classical terms $t/2+1/4$ in \eqref{TA610} takes into 
account that the corresponding domainwall not only changes the vacuum on the brane 
($\sigma=\pm 1$), but also the value of the Ramond-Ramond four and six form flux, $N_4$ 
and $N_6$, through the corresponding cycles of the Calabi-Yau manifold. We have 
\cite{Walcher:2006rs}
\begin{equation}
\eqlabel{allfluxes}
\T_A = \calw_{(N_4+1,N_6;+)} - \calw_{(N_4,N_6;-)} =
t-\bigl[\calw_{(N_4,N_6;-)}-\calw_{(N_4,N_6;+)}\bigr]~.
\end{equation}
The second equality follows from the fact that the domainwall mediating between $N_4$
and $N_4+1$ has tension equal to $t$. The large volume monodromy $t\to t+1$ acts on
the vacua as follows
\begin{equation}
\eqlabel{lvmono}
\begin{split}
(N_4,N_6;+) &\to (N_4,N_6+N_4;-)~, \\
(N_4,N_6;-) &\to (N_4+1,N_6+N_4;+)~.
\end{split}
\end{equation}
It is not hard to see that \eqref{TA610} is the only form consistent with these 
constraints. We emphasize that the monodromy \eqref{lvmono} as well
as the ``one-loop'' correction $\frac 14$ in \eqref{TA610} have not yet been derived 
from first principles, i.e. couplings of D-branes to Ramond-Ramond flux.

\subsection{\texorpdfstring{$X^{(8)}$ in the A-model}{X(8) in the A-model}}
\label{X8Amodel}

When all $h_i$ are even, as happens in our examples for $X^{(8)}$, the topological type
of $L^{(k)}_{[M]}$ cannot be determined straightforwardly by the previous argument, and in 
fact strongly depends on $M$. The problem was studied in a different context in 
\cite{rrw}. It is not hard to see that in the present case we have the following types
\begin{equation}
\eqlabel{cases}
\begin{split}
L_{[0,0,0,0,0]}&=\{y_1^8+y_2^8+y_3^8+y_4^8+y_5^2=0\} \cong \emptyset~, \\
L_{[0,0,0,0,1]}&=\{y_1^8+y_2^8+y_3^8+y_4^8-y_5^2=0\} \cong \RP^3\cup\RP^3 ~,\\
L_{[0,0,0,1,0]}&=\{y_1^8+y_2^8+y_3^8-y_4^8+y_5^2=0\} \cong S^3~ ,\\
L_{[0,0,0,1,1]}&=\{y_1^8+y_2^8+y_3^8-y_4^8-y_5^2=0\} \cong (S^1\times S^2)/\zet_2 ~,\\
L_{[0,0,1,1,0]}&=\{y_1^8+y_2^8-y_3^8-y_4^8+y_5^2=0\} \cong (S^1\times S^2)/\zet'_2~ .\\
\end{split}
\end{equation}
The distinction between the last two lines is in the action of $\zet_2$ on $S^1\times S^2$.
For $[M]=[0,0,0,1,1]$, it acts by an anti-podal map on $S^2$, and as inversion of $S^1$.
The Lagrangian $L=L_{[0,0,0,1,1]}$ in this case can be thought of as an $S^1$ bundle over 
$\RP^2$, with $H_1(L;\zet)=\zet\times\zet_2$, and $H_2(X,L;\zet)=\zet\times\zet$.
For $[M]=[0,0,1,1,0]$, the residual $\zet'_2$ acts by a half-shift on the $S^1$ and by
inversion of the longitudinal direction on $S^2$. The Lagrangian in this case is an 
$S^2$ bundle over $\RP^1\cong S^1$, with $H_1(L;\zet)=\zet$ and $H_2(X,L;\zet)=\zet\times\zet$.

In both cases, the Lagrangian contains a real one-cycle, namely the first Betti number $b_1(L)=1$.
As is well-known, this means that the $\caln=1$ worldvolume theory contains a chiral 
multiplet whose vev measures displacement of the Lagrangian away from its original position 
at the fixed point locus, as well as the continuous Wilson line around the corresponding 
one-cycle. As is equally well-known, this chiral multiplet is massless in the large volume
limit, but can gain a mass by worldsheet disk instantons \cite{kklm} (\ie, it could be an obstructed
deformation in the mathematical langage, \cite{fooo}) with boundary in the corresponding 
one-cycle. Using mirror symmetry, we can study the corresponding deformation problem using
classical methods. Preliminary computations on the objects mirror to the above Lagrangians
(see below) indicate that their modulus in fact remains massless even away from large volume 
\cite{unpub}. It would be interesting to find evidence for this vanishing of worldsheet 
instanton contribution directly in the A-model.

Another interesting case is the second line in \eqref{cases}, $L_{[0,0,0,0,1]}\cong
\RP^3\cup\RP^3$. Here, $H_1(L;\zet)=\zet_2\times\zet_2$, and $H_2(X,L;\zet)=\zet\times\zet_2$.
Thus, the fixed point locus actually consists of two components that can be wrapped 
{\it independently}. As we will see below, the two components are actually homologous to each
other, so that the worldvolume theory of a D-brane in this class has four vacua,
labelled by the $\RP^3$ component it is wrapped on, and the choice of discrete Wilson line
on the corresponding $\RP^3$. We will denote these moduli by $(\xi,\sigma)$, with $\xi,\sigma=
\pm 1$. 

\begin{figure}
\psfrag{A}[cc][][1]{$a)$}
\psfrag{B}[cc][][1]{$b)$}
\psfrag{C}[cc][][1]{$c)$}

\psfrag{a}[cc][][0.7]{$\sigma$}
\psfrag{b}[cc][][0.7]{$\xi$}
\psfrag{c}[cc][][0.7]{$\widetilde{\T_A}$}
\psfrag{d}[cc][][0.7]{$\T_A$}
\psfrag{e}[cc][][0.7]{$\R\P^3$}
\psfrag{f}[cc][][0.7]{$\R\P^3$}
\psfrag{g}[cc][][0.7]{$\Sigma$}

\begin{center}
\includegraphics[scale=0.3]{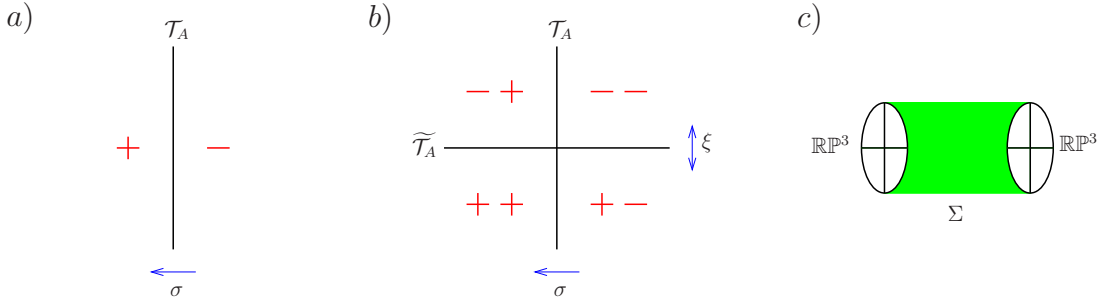}
\end{center}
\caption{Illustration of the vacua of the worldvolume gauge theory of a D6-brane on 
a) $L_{[M]}^{(6)}$ and $L_{[M]}^{(10)}$, b) on $L_{[0,0,0,0,1]}^{(8)}$. 
c) Illustration of the 4-chain $\Sigma$ separating the two $\R\P^3$ components of $L_{[0,0,0,0,1]}^{(8)}$.}
\label{modelsfig1}
\end{figure}

We illustrate the corresponding domainwalls in figure \ref{modelsfig1}.
First, we have the domainwall interpolating between the different Wilson lines on a fixed
Lagrangian. For symmetry reasons, the tension does not depend which $\RP^3$ component we
are talking about, and will be identical in structure to that on the quintic, $X^{(6)}$ and 
$X^{(10)}$, see \eqref{TA610}. In addition, we have the possibility of interpolating between 
the two $\RP^3$'s. This is realized geometrically as a D-brane partially wrapped on an 
appropriate four-chain, as illustrated in figure \ref{modelsfig1}c, with remaining directions 
extended in space-time. On dimensional grounds, the tension of this domainwall must scale as $t^2$ 
as $t\to\ii\infty$. In fact, one may see that by complex conjugation, we can complete the 
four-chain to a four-cycle, where a D6-brane wrapped on this four-cycle changes the 
two-form flux $N_2$ by one unit. The tension of this domainwall, $\Pi_4$,
is nothing but the (quantum corrected) closed string period of the four-cycle. From 
closed string mirror symmetry \cite{cdgp,Klemm:1992tx}, we know that $\Pi_4$ has at large 
volume an expansion of the form
\begin{equation}
\eqlabel{Pi4exp}
\Pi_4 = \del_t\calf = -\kappa \frac{t^2}{2} + a t + b + \frac{1}{4\pi^2} \sum_{d} 
\tilde N_d d q^d~,
\end{equation}
where $\calf$ is the genus zero prepotential. Here $\tilde N_d$ are the closed string 
Gromov-Witten invariants, $\kappa$ is the classical triple intersection, and $b$ is 
related to the second Chern class of the Calabi-Yau. The number $a$ is slightly
ambiguous, but can be constrained by requiring integrality of monodromy or by related 
considerations of D-brane charge quantization. For $X^{(8)}$, we use the values 
\cite{Klemm:1992tx}
\begin{equation}
\eqlabel{values}
\kappa=2\,, \qquad a = -3 \,, \qquad b = \frac{11}{6}~.
\end{equation}
Thus, under large volume monodromy $t\to t+1$, 
\begin{equation}
\Pi_4 \to \Pi_4 +a' t  + b'~,
\end{equation}
where $a'=-\kappa=-2$ and $b'=a-\frac\kappa2=-4$. 
We can now repeat the same steps that led to \eqref{TA610}, taking into
account also the 2-form flux. We find that the only way to obtain a consistent
solution to the monodromy constraints is that $t\to t+1$ acts on the vacua
by
\begin{equation}
\eqlabel{allmono}
\begin{split}
(N_2,N_4,N_6;-,+) &\to (N_2,N_4+a' N_2,N_6+N_4+b' N_2;-,-)~, \\
(N_2,N_4,N_6;-,-) &\to (N_2,N_4+a' N_2+1,N_6+N_4+b' N_2;-,+)~,\\
(N_2,N_4,N_6;+,+) &\to (N_2,N_4+a' N_2+ \frac{a'}2,N_6+N_4+b' N_2+\frac{b'}2;+,-)~, \\
(N_2,N_4,N_6;+,-) &\to (N_2,N_4+a' N_2+\frac{a'}2+1,N_6+N_4+b' N_2+\frac{b'}2;+,+)~,\\
\end{split}
\end{equation}
and that with
\begin{equation}
\eqlabel{TA8}
\begin{split}
\T_A &= \calw_{(N_2,N_4+1,N_6;\xi,+)} - \calw_{(N_2,N_4,N_6;\xi,-)} =
t-\bigl(\calw_{(N_2,N_4,N_6;\xi,-)}-\calw_{(N_2,N_4,N_6;\xi,+)}\bigr)~, \\
\widetilde{\T_A} &= \calw_{(N_2+1,N_4,N_6;-,\sigma)} - \calw_{(N_2,N_4,N_6;+,\sigma)}
= \Pi_4 -  \bigl(\calw_{(N_2,N_4,N_6;+,\sigma)}-\calw_{(N_2,N_4,N_6;-,\sigma)}\bigr)~,
\end{split}
\end{equation}
we must have an expansion of the form \eqref{TA610} for $\T_A$ and
\begin{equation}
\eqlabel{must}
\widetilde{\T_A} = \frac{\Pi_4}2~,
\end{equation}
with no further corrections. Note that the existence of a solution, and in particular
the integrality of the monodromy of $\widetilde{\T_A}$ depends on the fact that
$a'$ and $b'$ are even integers. In section \ref{monodromy}, we will check that 
the monodromies around the other singular points in moduli space are also integral.

\subsection{Comparison with B-model}

At the Fermat point $\psi=0$, the polynomials in \eqref{modelseq1}, viewed as Landau-Ginzburg
potentials, $W=W^{(k)}$, admit the following set of matrix factorizations 
\begin{equation}
\eqlabel{resc}
Q = \sum_{i=1}^5 \frac{1}{\sqrt{h_i}} 
\bigl(x_i^{l_i}\eta_i + x_i^{h_i-l_i} \bar\eta_i\bigr)~,
\end{equation}
where $\{\eta_i,\bar\eta_j\} = \delta_{ij}$ are matrices representing a Clifford algebra, 
and $0< l_i< h_i$ are a set of integer parameters. Namely,
\begin{equation}
\bigl(Q\bigr)^2 = \sum x_i^{h_i} / h_i = W |_{\psi=0}~.
\end{equation}
The factorizations in \eqref{resc} provide the Landau-Ginzburg description of the so-called
Cardy or Recknagel-Schomerus boundary states \cite{resc} of the associated Gepner 
model. More precisely, we are interested in B-branes in the mirror model, which involves an 
orbifold of \eqref{modelseq1} by the Greene-Plesser orbifold group $\hat G=\ker\bigl
(\prod\zet_{h_i}\to \zet_k\bigr)$. This means that we have to equip the linear space 
underlying $Q$ with an action of $\hat G$ such that $Q$ is equivariant with respect
to the action of $\hat G$ on the $x_i$.

As shown in \cite{bhhw}, the boundary states/matrix factorizations that provide the 
Landau-Ginzburg description of the real slices of the A-model hypersurfaces arise from
the labels $l_i\approx h_i/2$ for $i=1,\ldots, 5$. We will momentarily describe this correspondence. 
But before that, we ought to note that the factorizations \eqref{resc} in which $l_i=h_i/2$
for all $i$ with odd $\nu_i$ ($\equiv k/h_i$) are {\it reducible}. This is because
\begin{equation}
\eqlabel{AA}
A = \prod_{i,\nu_i\;{\rm odd}} (\eta_i-\bar \eta_i)~,
\end{equation}
is a non-trivial degree zero element of the cohomology of $Q$, of square $A^2\sim {\rm 
id}$. As first discussed in the Gepner model context in \cite{brsc,fsw}, we can then 
split $Q$ into the eigenspaces of $A$, as in 
\begin{equation}
Q^\pm = P^\pm Q P^\pm \, ,
\end{equation}
where $P^\pm \sim \frac{1\pm A}2$. We will denote the elementary matrix factorization,
equipped with the corresponding representation of $\hat G$, by
\begin{equation}
\eqlabel{splitted}
Q^\zeta_{[{\bf m}]}~,
\end{equation}
where $[{\bf m}]\in \bigl(\hat G\bigr)^*=\bigl(\prod\zet_{h_i}\bigr)/\zet_k$,
and $\zeta=\pm 1$ is the eigenvalue of $A$.

The correspondence derived in \cite{bhhw} is that the real slice $L^{(k)}_{[M]}$ of 
an {\it even-degree} hypersurface with respect to the involution \eqref{modelseq2}, 
is represented, {\it at the level of topological charges}, by the following linear 
combination of tensor product states:
\begin{equation}
\eqlabel{general}
\bigl[L_{[{M}]}\bigr] = \frac 12 \bigl[Q_{[{\bf m}^+]}\bigr] - \frac 12 
\bigl[Q_{[{\bf m}^-]}\bigr]~.
\end{equation}
Here ${\bf m}^\pm = (m_1^\pm,\ldots,m_5^\pm)$, and the $m_i^\pm$ are related to the 
$M_i$ as follows: For $M_i$ even, $m_i^+= m_i^- = M_i/2$, and for $M_i$ odd, 
$m_i^\pm = (m_i\pm 1)/2$. The $l_i$ labels (\cf, \eqref{resc}) are determined
as follows: For $h_i$ even, $l_i=h_i/2$. For $h_i$ odd, and $M_i$ even, $l_i=(h_i-1)/2$ 
in the first summand, and $l_i=(h_i+1)/2$ in the second summand. For $h_i$ odd, and
$M_i$ odd, $l_i=(h_i+1)/2$ in the first summand, and $l_i=(h_i-1)/2$ in the second
summand.

The relation \eqref{general} was obtained in \cite{bhhw} by comparing, via the gauged 
linear sigma model, the topological charges of {\it orientifold planes} associated with
A-type parity and complex conjugation \eqref{modelseq2} in large volume and in the 
Landau-Ginzburg phase. Our goal in this paper is however to obtain more refined 
information than just the topological charges, for which we need to lift \eqref{general} 
(at least) to the holomorphic sector. We have no principled way of doing this at the 
moment, however in certain cases we can make a plausible proposal based on the following 
set of observations.\footnote{For {\it odd degree} hypersurfaces, such as the 
quintic, we have only one term on the RHS of \eqref{general}. The lift to the
holomorphic sector is then more natural, and supported by a lot of evidence.}

At large volume, the fixed point set of the anti-holomorphic involution is a 
{\it special} Lagrangian submanifold, \ie, it is conformally invariant (in one-loop
sigma-model expansion) and preserves space-time supersymmetry, in addition to 
preserving A-type worldsheet supersymmetry. An equivalent statement should hold
at the Landau-Ginzburg point, since to get there we only need to vary the K\"ahler
moduli. In general, the ($\caln=1$) spacetime supersymmetry preserved by an equivariant 
matrix factorization $Q_{[{\bf m}]}$ can be measured by the phase of the ($\caln=2$) 
central charge, which (for fixed $l_i$) varies $\propto \sum_{i=1}^5 \frac{m_i}{h_i}$. 
It is not hard to see that in most cases, the two summands in \eqref{general} in general
preserve different supersymmetry. This means that the supersymmetric D-brane
corresponding to the real hypersurface must in general be some bound state of the
above components.

Let us illustrate this for the real slices of $X^{(8)}$. Evaluating \eqref{general}
(and taking into account that the irreducible factorizations from \eqref{splitted} 
have the same topological charges) gives (\cf, \eqref{cases}),
\begin{equation}
\eqlabel{virtual}
\begin{array}{c|c|c}
\text{Langrangian}& \text{topology}  & \text{matrix factorizations} \\
\hline
L_{[0,0,0,0,0]} &  \emptyset & \bigl[Q_{[0,0,0,0,0]}^+\bigr] -\bigl[Q_{[0,0,0,0,0]}^+\bigr]=0 \\
L_{[0,0,0,0,1]} & \RP^3\cup\RP^3 & \bigl[Q_{[0,0,0,0,0]}\bigr] = 
\bigl[Q_{[0,0,0,0,0]}^+\bigr] +\bigl[Q_{[0,0,0,0,0]}^-\bigr]\\
L_{[0,0,0,1,0]} & S^3 & \bigl[Q_{[0,0,0,0,0]}^+\bigr] - \bigl[Q_{[0,1,0,0,0]}^+\bigr] \\
L_{[0,0,0,1,1]} & (S^1\times S^2)/\zet_2 &  \bigl[Q_{[0,0,0,0,0]}^+\bigr] +
\bigl[Q_{[0,1,0,0,0]}^+\bigr]  \\
L_{[0,0,1,1,0]} & (S^1\times S^2)/\zet_2' & \bigl[Q_{[0,0,0,0,0]}^+\bigr] -
\bigl[Q_{[0,1,1,0,0]}^+\bigr]
\end{array}
\end{equation}
The first two lines are very obvious cases: $L_{[0,0,0,0,0]}$ is empty, with vanishing
boundary state. $L_{[0,0,0,0,1]}$ geometrically splits into two $\RP^3$ components,
which we might tentatively identify with $Q^+_{[0,0,0,0,0]}$ and $Q^-_{[0,0,0,0,0]}$. 
(The correct dictionary must ultimately include the Wilson line degree of freedom, and 
is somewhat different, see eq.\ \eqref{finally}.) We propose that this 
identification holds at the holomorphic level, and probably also at the level of 
superconformal boundary states.

The situation for the other real slices (including those of $X^{(6)}$ and $X^{(10)}$)
is less clear cut. As mentioned above, the Lagrangians can at best correspond to a 
bound state of the two components in \eqref{general}, and at worst might not be 
continuously connected to the split form of \eqref{virtual}. Nevertheless, our
present observations and the calculations in the following sections suggest that 
the correspondence \eqref{virtual} can indeed be lifted to the holomorphic level.

To study this additional evidence, we need to present the deformation theory
of the matrix factorizations $Q$ of \eqref{resc} with $l_i=[h_i/2]$ as we vary the 
complex structure parameter away from $\psi=0$. This is a rather straightforward 
exercise.

For $Y^{(6)}$, we find that $Q$ deforms in a unique way (up to gauge transformation), 
given explicitly by
\begin{equation}
\eqlabel{mf6}
Q (\psi) = \sum \frac{1}{\sqrt{6}}\bigl(x_i^3 \eta_i + x_i^3 \bar\eta_i\bigr) + 
\frac{1}{\sqrt{3}}\bigl(x_5\eta_5+x_5^2\bar\eta_5\bigr)
- \sqrt{3}\psi x_1x_2x_3x_4 \bar\eta_5~.
\end{equation}
This deformation commutes with $A$ from \eqref{AA}. Therefore, by splitting as
in \eqref{splitted}, we obtain two families $Q^\zeta(\psi)$ (with $\zeta=\pm 1$)
of matrix factorizations. We expect that the $\RP^3$ special Lagrangians should 
correspond to an appropriate bound state of those with different ${[\bf m]}$ label,
but identical $\zeta$-label. The latter should correspond to the discrete Wilson
line on $\RP^3$. Hence, we identify
\begin{equation}
\eqlabel{sigmazeta}
\sigma = \zeta ~.
\end{equation}

For $Y^{(8)}$, we find two {\it inequivalent} ways of deforming the factorization away
from $\psi=0$. We will denote those matrix factorizations as $Q(\psi,\mu)$, where the
additional label $\mu=\pm 1$:
\begin{equation}
\eqlabel{mf8}
\begin{split}
Q(\psi,+) &= \sum \frac{1}{\sqrt{8}}\bigl(x_i^4 \eta_i + x_i^4\bar\eta_i\bigr)
+\frac{1}{\sqrt{2}} \bigl((x_5\eta_5 + x_5\bar\eta_5\bigr)- \sqrt{2}\psi x_1x_2x_3x_4 \eta_5~, \\
Q(\psi,-) &= \sum \frac{1}{\sqrt{8}}\bigl(x_i^4 \eta_i + x_i^4\bar\eta_i\bigr)
+\frac{1}{\sqrt{2}} \bigl((x_5\eta_5 + x_5\bar\eta_5\bigr)- \sqrt{2}\psi x_1x_2x_3x_4 \bar\eta_5~ .
\end{split}
\end{equation}
Again, the deformation commutes with $A$. For $L_{[0,0,0,0,1]}$, this means that we 
obtain in total {\it four} families of matrix factorizations, naturally organized in 
two sets of two. Namely, we mind to the labels $\langle\mu,\zeta\rangle$, where $\zeta$ is the
eigenvalue of $A$, and $\mu$ distinguishes the two lines in \eqref{mf8}. We propose 
that those correspond to the four vacua that we identified in subsection 
\ref{X8Amodel} above. We emphasize at this stage that we still allow for a non-trivial 
transformation between the discrete A-model labels $(\xi,\sigma)$ and the B-model 
labels $\langle\mu,\zeta\rangle$. We will determine this transformation after analytic 
continuation of domainwall tensions in section \ref{monodromy}.

For $L_{[0,0,0,1,1]}\cong (S^1\times S^2)/\zet_2$, we propose to identify the 
$\mu=\pm 1$ label from \eqref{mf8} with the two vacua associated with the discrete 
$\zet_2$ factor in $H_1(L_{[0,0,0,1,1]})=\zet\times\zet_2$ (see paragraph below 
\eqref{cases}). (The free factor in $H_1(L)$ (for $L=L_{[0,0,0,1,1]}$ and $L_{[0,0,1,1,0]}$) 
is associated at large volume with a continuous modulus, displacing the Lagrangian
away from the fixed locus of the anti-holomorphic involution. As mentioned
above, there are indications \cite{unpub} that this open string modulus is in fact 
unobstructed, so should decouple from the superpotential computations.)

As on the quintic \cite{Walcher:2006rs,strings}, the two-fold way of deforming away 
from $\psi=0$ is accompagnied by the appearance, at $\psi=0$, of an additional
massless field in the open string spectrum. Also, the tension of the domainwall
between the $\langle+,\zeta\rangle$ and the $\langle-,\zeta\rangle$ vacua should 
vanish at $\psi=0$. This will be our way to complete the identification of the 
four vacua in A- and B-model.

Finally, for $Y^{(10)}$, the situation is somewhat in between that of $Y^{(6)}$
and that of $Y^{(8)}$. The main difference to $Y^{(6)}$ is that the tensor product
factorization has an infinitesimal modulus (degree $1$ cohomology element) $\Psi$, 
the main difference to $Y^{(8)}$ is that $\Psi$ satisfies $\{A,\Psi\}=0$ instead
of $[A,\Psi]=0$, where $A$ is from \eqref{AA}. Without delving into details, the 
consequence is that the factorizations $Q^\pm$ from \eqref{splitted} deform in a 
unique way, which can be obtained by splitting the deformed factorization
\begin{multline}
\eqlabel{mf10}
Q(\psi) = \sum \frac{1}{\sqrt{10}}\bigl(x_i^5\eta_i + x_i^5 \bar\eta_i\bigl) + 
\frac{1}{\sqrt{2}} \bigl(x_4\eta_4+x_4\bar \eta_4\bigr) +
\frac{1}{\sqrt{5}}
\bigl(x_5^2 \eta_5+ x_5^3 \bar\eta_5\bigr)\\
-\frac{\psi}{\sqrt{2}} x_1x_2x_3 (\eta_4+\bar\eta_4) x_5
- \frac{\psi^2}{2} \sqrt{5} x_1^2x_2^2x_3^2 \bar\eta_5 ~,
\end{multline}
in eigenspaces of $A$. Again, we identify the eigenvalue of $A$ with the discrete
Wilson line on $\RP^3\cong L_{[M]}^{(10)}$, as in \eqref{sigmazeta}

Since the factorizations on $Y^{(6)}$ and $Y^{(10)}$ deform in a unique way, there is no
additional massless open string, and we expect no tensionless domainwall at $\psi=0$.
We will confirm this in section \ref{monodromy}.

Before closing this section, we note another property (valid for all three $k$'s) of 
the factorizations around the Fermat point $\psi=0$. This is a special point in moduli 
space in which the hypersurfaces $Y^{(k)}$ gain an additional $\zet_k$ automorphism 
multiplying one of the weight-one variables by a phase. Put differently, the 
{\it monodromy around the Gepner point} $\psi\to\ee^{2\pi\ii/k}\psi$ can 
be undone by rotating $x_1\to \ee^{-2\pi\ii/k} x_1$. At the level of 
matrix factorizations, this monodromy has to be accompagnied by conjugating 
$Q$ with a representation of the $\zet_k$ symmetry group of the corresponding 
minimal model $x_1^k$. It is not hard to see that the matrix $A$ from
\eqref{AA} {\it picks a sign} under this symmetry. Thus, we conclude that
{\it monodromy around the Gepner point exchanges the vacua labelled by $\zeta=\pm 1$}.%
\footnote{To complete this, note that for $Y^{(8)}$, Gepner monodromy does not affect the 
$\mu$-label, as can also be seen from \eqref{mf8}.} This is another clue that we will 
pick up in our monodromy discussion in section \ref{monodromy}.

\section{Inhomogeneous Picard-Fuchs equations}
\label{picardfuchs}

We now wish to determine the non-perturbative $\alpha'$ (disk instanton) corrections to
the classical approximations to the domainwall tensions \eqref{TA610}, \eqref{TA8} between the
various vacua on our branes. The basic strategy is as follows.

For each A-model domainwall, we identified a (virtual linear combination of) matrix factorization 
$Q$ describing the D-brane configuration in the mirror B-model. The algebraic
second Chern class of this matrix factorization can be represented by a homologically trivial
codimension-2 algebraic cycle $C$ (in other words, an integral linear combination of 
holomorphic curves)
\begin{equation}
c_2(Q) = [C] \in {\rm CH}_{\rm hom}^2(Y)~,
\end{equation}
which we will explicitly compute from the matrix factorization. There then exists a 
three-chain $\Gamma$ of boundary $C$, well-defined up to closed three-cycle $\Gamma^c\in 
H_3(Y;\zet)$. The domainwall tension is computed by the integral over $\Gamma$ 
\begin{equation}
\eqlabel{chainint}
\calt_B(z) = \int_\Gamma \hat \Omega(z)~,
\end{equation}
where $\hat\Omega(z)$ is the appropriately normalized holomorphic three-form of the B-model
geometry (see below).

Explicitly, we will have relations of the form
\begin{equation}
\eqlabel{mirrormap}
\calt_B(z(t))/\varpi_0(z(t)) = \calt_A(t)~,
\end{equation}
where the mirror map consists of the relation $z=z(t)$ between A- and B-model variables, and
the normalization of the holomorphic three-form $\hat\Omega(z)\to \hat\Omega(z)/\varpi_0(z)$. 
As is well-known, this data can be obtained by solving the homogenous Picard-Fuchs equation 
satisfied by the B-model periods. The Picard-Fuchs operators of our three models are:
\beq\eqlabel{modelseq3}
\begin{split}
\PF^{(6)}&:=\th^4-2^4 3^6z(1/6+\th)(1/3+\th)(2/3+\th)(5/6+\th)~,\\
\PF^{(8)}&:=\th^4-2^{16}z(1/8+\th)(3/8+\th)(5/8+\th)(7/8+\th)~,\\
\PF^{(10)}&:=\th^4-5^5 2^8 z(1/10+\th)(3/10+\th)(7/10+\th)(9/10+\th)~,
\end{split}
\eq
with $\th=z\D_z$, and $z\sim \psi^{-k}$. Namely, $\varpi_0(z)$ is the unique solution with 
power series behavior at $z=0$, and if $\varpi_1(z)\sim \varpi_0(z) \log(z)$ is the solution 
with a single logarithm, we have
\begin{equation}
t(z) = \frac{\varpi_1(z)}{\varpi_0(z)}~.
\end{equation}

To calculate the chain integral in \eqref{chainint}, we exploit that it satisfies
an inhomogeneous version of the Picard-Fuchs equation,
\beq\eqlabel{modelseq4}
\PF^{(k)} \T_B= \frac{c^{(k)}}{16} \sqrt{z}~.
\eq
The central part of our computation is the determination of the parameters $c^{(k)}$
for each of our domainwalls.

\subsection{From matrix factorizations to curves}

Given the matrix factorizations, we obtain the curves representing the algebraic
second Chern classes by the algorithm described in \cite{Morrison:2007bm} for the
quintic. This can be viewed as an application of the homological 
Calabi-Yau/Landau-Ginzburg correspondence \cite{orlov2,hhp} together with elementary
methods for the computation of Chern classes. The only point on which we will be
slightly less rigorous than in \cite{Morrison:2007bm} is that we will perform
our computation as if we were pretending to be working on the hypersurfaces 
\eqref{modelseq1} in weighted projective space, without orbifold. In actuality, 
however, everything is taking place on the B-model side, \ie, the underlying manifolds 
are indeed $Y^{(k)}$, {\it after} quotienting by $G=\hat G/\zet_k$.

After some algebra, we find that the relevant part of the second Chern classes 
of the matrix factorizations from eqs.\ \eqref{mf6}, \eqref{mf8}, \eqref{mf10} 
can be represented with the following set of curves.
\begin{equation}
\eqlabel{curves}
\begin{split}
k=6: &\; C^{(6)}_\zeta = \{ x_1 +(\alpha^{(6)})^{\zeta} x_2 =0, x_3 + \alpha^{(6)} x_4 = 0, 
x_5^2 -3\psi x_1x_2x_3x_4 =0 \} \\
k=8: &\;  C^{(8)}_{\langle\mu,\zeta\rangle} = 
\begin{cases}\{ x_1 +  (\alpha^{(8)})^{\zeta} x_2 =0, x_3+\alpha^{(8)} x_4=0, x_5 = 0 \} 
&\!\!\!\! \mu = +1 \\
\{ x_1+(\alpha^{(8)})^{\zeta}  x_2 =0, x_3+\alpha^{(8)}x_4 = 0 , x_5-2\psi x_1x_2x_3x_4 = 0\}
&\!\!\!\! \mu = -1
\end{cases} \\
k=10: &\;  C^{(10)}_\zeta = \{ x_1 + (\alpha^{(10)})^\zeta x_2 =0 , x_3^5 +(\alpha^{(10)})^{5} 
\sqrt{5}\bigl(x_5-\psi x_1x_2x_3x_4\bigr) = 0, \\
&\qquad\qquad\qquad\qquad\qquad \qquad\qquad\qquad\qquad\qquad\qquad 
2 x_4^3 - 5 \psi^2 x_1^2x_2^2x_3^2 = 0 \}~.
\end{split}
\end{equation}

Let us explain our notation and the precise meaning of those equations. First of all, 
$\alpha^{(k)}$ are phases which fulfill $(\alpha^{(k)})^k=-1$. One then
easily checks that the curves actually lie on the corresponding hypersurface,
\ie, $W^{(k)}=0$ identically whenever a group of three equations from \eqref{curves}
are satisfied. The subscript $\zeta$ comes from the eigenvalue of $A$ used to split the 
reducible $Q$ from \eqref{resc} in two components (for the geometric meaning of
$\zeta$, see two paragraphs below). Since the second Chern class does not depend 
on the representation of $\hat G$, we have omitted the $[\bf m]$ label.

Next, we notice that the curves in \eqref{curves} are not invariant under the
orbifold group $G$. Instead, $G$ maps each curve to a similar one with 
different choices of $k$-th roots of $-1$ in the corresponding equation, and the
second Chern class should be thought of as this orbit of curves in $Y^{(k)}$. However, 
note that in each case, a certain subgroup $G_{\rm fix}\subset G$ does 
leave the curve invariant. This will lead to an additional normalization factor in 
our calculations in the next subsection. We identify the respective subgroups in 
table \ref{invgps}. 

\begin{table}
\begin{center}
\begin{tabular}{c|c|c}
$k$ & $G_{\rm fix}$ & generator \\\hline
6 &  $\zet_6$ & $(\gamma,\gamma,\gamma^{-1},\gamma^{-1},1)$ \\
8 &  $\zet_4$ & $(\gamma,\gamma,\gamma^{-1},\gamma^{-1},1)$  \\
10 &  $\zet_{10}$ & $(\gamma,\gamma,\gamma^8,1,1)$  
\end{tabular}
\caption{The curves from \eqref{curves} are invariant under certain subgroup of
the orbifold group $G$. For each $k$, $\gamma=\gamma_k=\exp(2\pi\ii/k)$.}
\label{invgps}
\end{center}
\end{table}

To explain the geometric role of the parameter $\zeta=\pm 1$ in \eqref{curves}, we 
consider the case $k=6$ (the discussion on the other two models is the same). 
The set of curves $\{x_1+\alpha x_2=0,x_3+\beta x_4=0,x_5^2 - 3\psi x_1x_2x_3x_4=0\}$,
where $\alpha$ and $\beta$ are arbitrary $6$-th roots of $-1$ all lie on $X^{(6)}$.
Those curves organize into two distinct orbits under the action of $G^{(k)}$,
precisely distinguished by $\zeta=\pm 1$. It is clear that the Gepner monodromy
$\psi\to\ee^{2\pi\ii/k}\psi$, $x_1\to \ee^{-2\pi\ii/k} x_1$ exchanges those
two orbits, $\zeta\to-\zeta$, just as we had noted it at the end of the previous
section.

\subsection{From curves to Picard-Fuchs}
\label{maincompu}

We derive the inhomogeneous Picard-Fuchs equation, \ie, the constants
$c^{(k)}$ in \eqref{modelseq4} by using the Griffiths-Dwork method.  We 
give the details of the computation in the appendix, and only summarize
the salient steps here. 

The normalization of the holomorphic three-form for which we will quote the
$c^{(k)}$ is given by 
\beq
\eqlabel{normalization}
\h\Omega=\frac{\Ord(G^{(k)})}{(2\pi i)^3}\p\Omega~,
\eq
where $\Omega$ is the standard residue from projective space (see below (\ref{Dworkeq0a})) 
and the $\Ord(G^{(k)})$ are the orders of the groups yielding the mirror manifold. We list 
those together with some other normalization data in table \ref{maintab1}. The choice
\eqref{normalization} is the normalization in which the regular integral period 
$\varpi_0(z)$ has a unit constant term, $\varpi_0(z)= 1 + \calo(z)$.

\begin{table}
\begin{center}
\begin{tabular}{c|cccc}
k&$\Upsilon^{(k)}$&$\Ord(G^{(k)})$&$\sqrt{z}$&$\rho^{(k)}$\\
\hline
$6$&$6^{-4}\p^{-3}$&$3\cdot 6^2$&$2\cdot6^{-3}\p^{-3}$&$3^2 \pi^{-4}$\\
$8$&$2^{-12}\p^{-6}$&$2\cdot 8^2$&$2^{-8}$&$2^3\pi^{-4}\p^{-2}$\\
$10$&$10^{-4}\p^{-8}$&$10^2$&$2^{-4}5^{-5/2}\p^{-5}$&$4\cdot 5^{1/2}\pi^{-4}\p^{-3}$\\
\end{tabular}
\caption{Parameters used in the main text for the hypersurfaces under consideration.}
\label{maintab1}
\end{center}
\end{table}

The implementation of the Griffiths-Dwork algorithm however is easier in a slightly
different normalization of the holomorphic three-form, in which the homogeneous Picard-Fuchs 
operator takes the form $\t\PF^{(k)}$ given in equations (\ref{Dworkeq1}), (\ref{Dworkeq2}) 
and (\ref{Dworkeq3}). They are related to the Picard-Fuchs operators $\PF^{(k)}$ given 
in equation (\ref{modelseq3}) by the relation
\beq
\PF^{(k)}=-\Upsilon^{(k)} \t\PF^{(k)}\frac{1}\psi~,
\eq
where the $\Upsilon^{(k)}$ are given in table \ref{maintab1}. 

Putting the normalization together, the statement that $\T_B$ given in \eqref{chainint}
(for a certain choice of curves and three-chain, explained momentarily) satisfies the 
inhomogeneous Picard-Fuchs equation given in (\ref{modelseq4}) translates to the identity
\beq
\eqlabel{mainclaimeq4}
\t\L^{(k)}\int_{T_\e(\G)}\t\Omega=-\frac{(2\pi i)^4}{\Upsilon^{(k)} \Ord(G^{(k)})}c^{(k)}\sqrt{z}~,
\eq
where $\t\Omega$ is defined in (\ref{Dworkeq0a}) and where we have used (\ref{Dworkeq0b}). 
If the functional form $\propto \sqrt{z}$ is correct, the $c^{(k)}$ 
are the only unknown parameters, and the claim reduces to determine the constants 
$c^{(k)}$ in (\ref{mainclaimeq4}). That is, we can solve for $c^{(k)}$:
\beq\label{mainclaimeq4a}
c^{(k)}=-\rho^{(k)}\t\L^{(k)}\int_{T_\e(\G)}\t\Omega~,
\eq
where the $\rho^{(k)}$ are given in table \ref{maintab1}.

In \eqref{mainclaimeq4}, $T_\e(\G)$ is a small tube around a three-chain suspended 
between the appropriate combination of curves. We are interested in the following
B-model integrals (the curves are defined in \eqref{curves}):
\begin{equation}
\eqlabel{combinations}
\begin{array}{c|cc}
k & \text{domainwall tension} & \text{curve combination} \\\hline
6 & \T_B  & C^{(6)}_+ - C^{(6)}_- \\
8 & \T_B  & C^{(8)}_{\langle+,+\rangle} - C^{(8)}_{\langle-,+\rangle} \\
8 & \widetilde{\T_B} & C^{(8)}_{\langle+,+\rangle} - C^{(8)}_{\langle-,-\rangle} \\
10 & \T_B & C^{(10)}_+ - C^{(10)}_-
\end{array}
\end{equation}
For $k=8$, we make the choice to denote by $\T_B$ the domainwall tension that vanishes 
at $\psi=0$, which completely specifies the 3-chain between the respective curves. 
All other domainwall tensions are only define modulo integral periods for the moment.

There are in principle two types of contributions to $\tilde\L^{(k)}\int_{T_\e(\G)}\tilde\Omega$, 
depending on whether $\tilde\L^{(k)}$ acts on the holomorphic three-form, or on the
(tube over the) three-chain. As on the quintic, it turns out that the latter
contribution always vanishes (see appendix). Thus, we just need to evaluate
\beq\label{mainclaimeq5}
\int_{T_\e(\G)}\t\L^{(k)}\t\Omega=\int_{T_\e(\D\G)}\t\beta^{(k)}~,
\eq
where $\t\beta^{(k)}$ are the exact parts of the inhomogeneous Picard-Fuchs equations 
given in equations (\ref{Dworkeq1a}), (\ref{Dworkeq2a}) and (\ref{Dworkeq3a}).
Thus the crucial integrals are those of $\t\beta^{(k)}$ over the tubes around the
curves $C^{(k)}_\zeta$ for $k\in\{6,10\}$ and $C^{(k)}_{\zeta\mu}$ for $k=8$.

The main property of the curves that allows the evaluation of these integrals is
their planarity. Namely, as on the quintic, the curves $C_\zeta^{(k)}$ and
$C^{(8)}_{\zeta\mu}$ are components of the intersection of the hypersurface with
an appropriately chosen plane $P^{(k)}$. Except for a small neighborhood of the 
intersection of the components of $P^{(k)}\cap Y^{(k)}$, the tube around the curves can 
be laid into the plane, where the meromorphic three-form $\tilde\beta^{(k)}$ vanishes
trivially. We give the remaining details of this calculation in the appendix. The 
results are the following:

\begin{equation}
\eqlabel{intresults}
\begin{split}
k=6:&\qquad \int_{T_\e(C^{(6)}_\zeta)} \t\beta^{(6)} = \zeta\;\frac{4}{3}\pi^2~,\\
k=8:& \qquad \int_{T_\e(C^{(8)}_{\zeta\mu})} \t\beta^{(8)} = \zeta\mu\; 3\pi^2\psi^2~,\\
k=10:& \qquad \int_{T_\e(C^{(10)}_\zeta)} \t \beta^{(10)} = \zeta\; \frac{16}{5} 
\sqrt{5} \pi^2 \psi^3 \,.
\end{split}
\end{equation}

Referring to \eqref{mainclaimeq4a} and (\ref{combinations}), this translates to the 
following values for the constants $c^{(k)}$ for each of our domainwalls:
\begin{equation}
\eqlabel{entries}
\begin{array}{c|cc}
k & \text{domainwall} & c^{(k)} \\\hline
6 & \T_B & \frac{24}{\pi^2} \\
8 & \T_B & \frac{48}{\pi^2} \\
8 &\widetilde{\T_B} & 0 \\
10 &\T_B & \frac{128}{\pi^2}  
\end{array}
\end{equation}
We now proceed to the explicit solution of the inhomogeneous Picard-Fuchs equation
\eqref{modelseq4}.

\section{Analytic Continuation}
\label{monodromy}

As on the quintic \cite{Walcher:2006rs}, it turns out that the constant $c^{(k)}$ that 
we computed in the previous section can almost uniquely be recovered by assuming an 
inhomogeneous term $\sim \sqrt{z}$ and requiring integrality of monodromy around the
various special points in moduli space. We will follow this route here, and connect
to the previous discussion at the end. We denote by $\tau^{(k)}(z)$ the solution 
of the corresponding fifth order operator $(2\theta-1)\PF^{(k)}$ with squareroot 
behaviour at $z=0$. Note that for simplicity, we will sometimes drop the $^{(k)}$ 
indices in the following, which quantity carries a $^{(k)}$ index should be clear 
from the context. 

\subsection{Solutions}

The solutions of the Picard-Fuchs equations of our hypersurfaces around $z=0$ can 
by obtained by the Frobenius method from the following hypergeometric generating
function
\begin{equation}
\eqlabel{generate}
\Pi^{(k)}(z;H) = \sum_{m=0}^\infty z^{m+H} \frac{\Gamma\bigl(k(m+H)+1\bigr)} 
{\prod_{i=1}^5 \Gamma\bigl(\nu_i(m+H)+1\bigr)}~.
\end{equation}
Namely, one checks that by expanding \eqref{generate} in powers of $H$,
\begin{equation}
\eqlabel{tobefair}
\Pi^{(k)}(z;H) = \sum_{j=0}^3 H^j \Pi_{2j}^{(k)}(z)  \bmod H^4~,
\end{equation}
the $\Pi_{2j}^{(k)}(z)$ satisfy $\PF^{(k)} \Pi_{2j}^{(k)}(z) = 0$. Quite remarkably, the 
additional solution of our inhomogeneous equation can be obtained by setting $H=1/2$ in 
\eqref{generate}
\begin{equation}
\eqlabel{setting}
\tau^{(k)}(z) = \Pi^{(k)}(z;1/2)~.
\end{equation}
It satisfies
\begin{equation}
\PF^{(k)}\tau^{(k)}(z) = \frac{\tilde c^{(k)}}{16} \sqrt{z}~,
\end{equation}
where
\begin{equation}
\eqlabel{where}
\tilde c^{(k)} = \frac{\Gamma\bigl(k/2+1\bigr)}
{\prod \Gamma\bigl(\nu_i/2+1\bigr)}~.
\end{equation} 

The radius of convergence of the series \eqref{setting} is, as for the closed
string periods, given by $|z|< R_* \equiv \prod \nu_i^{\nu_i}/k^k$. To analytically 
continue $\tau^{(k)}(z)$ to the rest of the moduli space, in particular the Gepner 
point $1/z=0$, we utilize the familiar integral representation
\begin{equation}
\eqlabel{representation}
\tau^{(k)}(z)= \frac{1}{2\pi\ii} \int 
\frac{\Gamma\bigl(\frac12-s\bigr)\Gamma\bigl(\frac12+s\bigr)\Gamma\bigl(ks+1\bigr)}
{\prod\Gamma\bigl(\nu_i s+1\bigr)}
\ee^{\ii\pi(s-\frac 12)} z^s~,
\end{equation}
where the integration contour runs straight up the imaginary axis. For $|z|< R_*$,
we close the contour on the positive real axis, pick up the poles at $s=m+\frac 12$
for $m=0,1,\ldots$, and recover \eqref{setting}. For $|z|> R^*$, we close on the negative
real axis, where we find poles of $\Gamma$-functions in the numerator at $s=-m-\frac 12$ 
and at $s=-\frac{m}{k}$ for $m=1,2,\ldots$. When $k$ is even, the second actually
encompass the former. In that case, however, we also have exactly one even weight
($\nu_5$ in our notation) so that there is also a pole in the denominator, and the
total pole at $s=-m-\frac 12$ is first order. To make progress, we separate the 
terms with half-integer power of $z$ from the terms with powers on the list of 
exponents of the homogeneous equation,
\begin{equation}
\eqlabel{progress}
\tau^{(k)}(z) = \tau_1^{(k)}(z) + \tau_2^{(k)}(z)~,
\end{equation}
where
\begin{equation}
\eqlabel{compare}
\begin{split}
\tau_1^{(k)}(z) & = \sum_{m=0}^\infty\frac{\nu_5}{k} (-1)^{\frac k2+\frac{\nu_5}2+1}
\frac{\Gamma\bigl(\nu_5 m + \frac{\nu_5}2\bigr)}
{\Gamma\bigl(k m + \frac k2\bigr)\prod_{i,\nu_i\;{\rm odd}}
\Gamma\bigl(1-\nu_i(m+\frac 12)\bigr)} \; z^{-m-\frac 12}~,\\
\tau_2^{(k)}(z) & = -\frac{\pi}{k} \sum_{m=1}^\infty 
\frac{\ee^{\ii\pi\bigl(m-\frac{m}k-\frac 12\bigr)}}{\cos \pi\frac mk} \;
\frac{1}{\Gamma(m)\prod\Gamma\bigl(1-\frac{\nu_i}k m\bigr)} \;
 z^{-m/k}~.
\end{split}
\end{equation}
In the sum for $\tau_2^{(k)}$, we have to exclude those $m$ for which $m/k$ 
is a half integer, since we have already attributed these terms to $\tau_1^{(k)}$.
On the other hand, all other terms for which $\nu_i m/k$ is integer can be 
trivially included. 

Our next task is to express $\tau_2^{(k)}$ in terms of the solutions of the
homogeneous equation. We use the set of solutions of \cite{Klemm:1992tx},
\begin{equation}
\eqlabel{overcomplete}
\vp_j^{(k)}(z) = -\frac{\pi}{k} \sum_{m=1}^\infty \frac{\ee^{\ii\pi(m-\frac mk)\;\ee^{2\pi\ii j m/k}}}
{\sin\pi\frac mk} \;
\frac{1}{\Gamma(m)\prod\Gamma\bigl(1-\frac{\nu_i}k m\bigr)}\;
z^{-m/k}~,
\end{equation}
for $j=0,\ldots,k-1$. Comparing \eqref{compare} with \eqref{overcomplete}, we find that
\begin{equation}
\eqlabel{notunique}
\tau_2^{(k)}(z) = \sum a_j \vp_j (z)~,
\end{equation}
provided the $a_j$ satisfy the equations
\begin{equation}
-\ii\; \tan\pi\frac mk = \sum_{j=0}^{k-1} a_j \ee^{2\pi\ii j m/k}~,
\end{equation}
for $m-0,1,\ldots,k-1$, and where the LHS is set to $0$ for $m=k/2$. Of course, 
the $a_j$ are not uniquely determined by \eqref{notunique}, because the $\varpi_j(z)$ 
satisfy some linear relations owing to the poles in the denominator of 
\eqref{overcomplete}. Following \cite{cdgp,Klemm:1992tx}, we will use the 
``period vector at the Gepner point''
\begin{equation}
\vp^{(k)} = \bigl(\vp_2, \vp_1,\vp_0,\vp_{k-1}\bigr)~,
\end{equation}
and write
\begin{equation}
\eqlabel{write}
\tau_2^{(k)} = \tilde a^{(k)} \cdot \vp^{(k)}~.
\end{equation}
We have the following results for the vectors $a^{(k)}$ and $\tilde a^{(k)}$.
\begin{equation}
\begin{array}{c|cc}
k & a^{(k)} & \t a^{(k)}  \\ \hline
 6 & \frac{1}{3}(0,-2,1,0,-1,2) & \frac 13 (2,-2,1,2)\\
 8 & \frac 14 (0,-3,2,-1,0,1,-2,3) & (1,-1,0,1) \\
10 & \frac 15 (0,-4,3,-2,1,0,-1,2,-3,4) & \frac 15 (2,-4,1,2) 
\end{array}
\end{equation}
Also from \cite{Klemm:1992tx}, we extract the analytic continuation matrices between the
Gepner basis $\vp^{(k)}$ and the large volume basis $\amalg^{(k)} = 
(\Pi_6,\Pi_4,\Pi_2,\Pi_0)$. (This basis is almost the one from \eqref{tobefair}. See
\cite{Klemm:1992tx} for precise definitions.) Namely,
\begin{equation}
\amalg^{(k)} = M^{(k)} \vp^{(k)}~,
\end{equation}
with
\begin{equation}
\eqlabel{bastrafo}
\begin{split}
M^{(6)}=
\begin{pmatrix}
0&-1&1&0\\
-1&0&3&2\\
\frac{1}{3}&\frac{1}{3}&-\frac{1}{3}&-\frac{1}{3}\\
0&0&1&0
\end{pmatrix}&\,,\qquad
M^{(8)}=
\begin{pmatrix}
0&-1&1&0\\
-1&0&3&2\\
\frac{1}{2}&\frac{1}{2}&-\frac{1}{2}&-\frac{1}{2}\\
0&0&1&0
\end{pmatrix}\,,\;\\
M^{(10)}=&
\begin{pmatrix}
0&-1&1&0\\
0&1&1&1\\
1&0&0&-1\\
0&0&1&0
\end{pmatrix} \,.
\end{split}
\end{equation}
Finally, we record the action of the Gepner monodromy 
\begin{equation}
\eqlabel{monoeq7}
z^{-\frac{1}{k}}\rightarrow \ee^{\frac{2\pi i}{k}}z^{-\frac{1}{k}}~,
\end{equation}
on the basis $\amalg^{(k)}$ (along the path leading to the above basis transformation).
We have $\amalg^{(k)}\to A^{(k)}\amalg^{(k)}$ with
\begin{equation}
\eqlabel{path}
\begin{split}
A^{(6)} = 
\begin{pmatrix}
-3&-1&-6&4\\
-3&1&3&3\\
1&0&1&-1\\
-1&0&0&1
\end{pmatrix} &\,,\qquad
A^{(8)} = 
\begin{pmatrix}
-3&-1&-4&4\\
-2&1&2&2\\
1&0&1&-1\\
-1&0&0&1
\end{pmatrix}
\,, \\
A^{(10)} = &
\begin{pmatrix}
-2&-1&-1&3\\
0&1&1&0\\
1&0&1&-1\\
-1&0&0&1
\end{pmatrix}
\,,
\end{split}
\end{equation}

\subsection{Monodromy}

We now have all expressions at our disposal to discuss the analytic continuation and 
monodromy properties of the open string periods. After the mirror map, our current 
ansatz for the B-model version of the domainwall tension with large volume
expansion \eqref{TA610} is\footnote{We use $A/B$ subscript to distinguish the large/small
volume basis, and leave the mirror map \eqref{mirrormap} implicit.}
\begin{equation}
\eqlabel{implicit}
\T_A^{(k)} = \frac{\Pi^{(k)}_2}{2} + \frac{\Pi^{(k)}_0}{4} + d \tau^{(k)}~,
\end{equation}
where $d$ is a constant. For $k=8$, the additional domainwall interpolating between
the two $\RP^3$ components of $L_{[0,0,0,0,1]}^{(8)}$ has tension
\begin{equation}
\eqlabel{explicit}
\widetilde{\T_A} = \frac{\Pi^{(8)}_4}2~.
\end{equation}
By construction, $\T_A$ and $\widetilde{\T_A}$ have integral large volume
monodromy as $t\to t+1$.
Namely,
\begin{equation}
\eqlabel{LVmono}
t\to t+1:\qquad
\begin{array}{rl}
\T_A  &\to \Pi_2 - \T_A \\
\widetilde{\T_A}  &\to \widetilde{\T_A} - \Pi_2 -2 \Pi_0~.
\end{array}
\end{equation}

Consider now the Gepner monodromy, using \eqref{bastrafo} and \eqref{path}. From the
splitting \eqref{progress}, we see that $\tau_1^{(k)}$ changes sign as we circle
$1/z = 0$ once. Since the squareroot is the hallmark of the inhomogeneity, we see 
that $\T_A$ must come back to minus itself, up to a solution of the homogeneous 
equation. To ensure that this solution is an integral period, we consider the 
behaviour of $\tau_2^{(k)}$. Combining \eqref{write} and \eqref{path}, we find
\begin{equation}
\tau_2^{(k)} \to - \tau_2^{(k)} + \Pi_6^{(k)}~.
\end{equation}
Since for each $k=6,8,10$, the classical part of the domainwall tension transforms as
\begin{equation}
\T_{A,{\it cl}} = \frac{\Pi_2}2 + \frac{\Pi_0}4 \to 
\frac{\Pi_6}4 +\Pi_2 -\T_{A,{\it cl}}~,
\end{equation}
we see that the minimal value of $d$ that guarantees integrality is
\begin{equation}
\eqlabel{magic}
d = -\frac 14~.
\end{equation}
The Gepner monodromy then acts as
\begin{equation}
\eqlabel{Gepmono}
z^{-1/k} \to \ee^{2\pi\ii/k} z^{-1/k}:\qquad
\begin{array}{rl}
\T_A & \to \Pi_2 - \T_A \\
\widetilde{\T_A} & \to \widetilde{\T_A} - \Pi_6 + \Pi_2 + \Pi_0~.
\end{array}
\end{equation}
Integrality in the last line is ensured by the evenness of the appropriate entries in 
\eqref{path}. A noteworthy consequence is the invariance of $\T_A$ under the conifold 
monodromy, resulting from combination of \eqref{LVmono} and \eqref{Gepmono}.

\subsection{Domainwall spectrum and final matching of vacua}

The domainwall tensions \eqref{implicit}, \eqref{explicit} satisfy the same inhomogeneous 
Picard-Fuchs equation that we had obtained in section \ref{picardfuchs}. Namely, 
with $\tilde c^{(k)}$ from \eqref{where} and $d=-1/4$ from \eqref{magic}, we find
\begin{equation} 
\frac 14 \tilde c^{(k)} = c^{(k)}~,
\end{equation}
where $c^{(k)}$ are the non-zero entries in \eqref{entries}. And $\widetilde{\T_A}$
satisfies the homogeneous equation by construction. 

The fact that $\T_A^{(k)}$ comes back to minus itself (up to an integral period) in 
\eqref{Gepmono} means that the corresponding brane vacua are exchanged under Gepner 
monodromy. This is precisely what we had noted at the end of section \ref{configs}.

Finally, we test for tensionless domainwalls at the Gepner point. Using \eqref{write}, we 
find that the leading behaviour of $\T_A$ around $1/z\to 0$ is given by
\begin{equation}
\eqlabel{leadingbehaviour}
\T_A \sim \sqrt{z} + \tilde b^{(k)} \cdot \amalg^{(k)}~,
\end{equation}
where $\tilde b^{(k)} = - \frac 14 (M^{(k)})^{-T} \tilde a^{(k)} + (0,0,\frac 12,\frac 14)$,
explicitly,
\begin{equation}
\eqlabel{btilde}
\begin{array}{c|c}
k  & \tilde b^{(k)} \\ \hline
6  & \frac 13(-2,-1,-3,4)\\
8  & (-1,-\frac 12,-1,2)\\
10 & \frac 15 (-2,-1,2,4)
\end{array}
\end{equation}
Thus, for $k=6,10$, the leading behaviour at $1/z=0$ is always dominated by a non-vanishing
closed string period, and we have no tensionless domainwall.\footnote{This does not exclude
the interesting possibility that there are tensionless domainwalls somewhere else in the 
moduli space.} This is exactly what we predicted in section \ref{configs} under the 
identification \eqref{sigmazeta}, and concludes our discussion for those two models.

We continue with $k=8$. First of all, we see from \eqref{btilde} that by combining $\T_A$ 
with $\widetilde{\T_A}$, we obtain a tensionless domainwall\footnote{This, as well as all 
remaining statements in this subsection, are understood modulo integral periods. Those
correspond to changing Ramond-Ramond flux, which is invisible on the brane. We also leave 
the mirror map implicit.} at the Gepner point, which is again precisely as predicted! In 
section \ref{picardfuchs}, we had denoted this vaninishing domainwall by $\T_B$, so we 
identify
\begin{equation}
\T_B = \T_A + \widetilde{\T_A}~.
\end{equation}
The other open string period from section \ref{picardfuchs} satisfies the homogeneous 
equation, so we have
\begin{equation}
\widetilde{\T_B} = \widetilde{\T_A}~.
\end{equation}

\begin{figure}
\psfrag{A}[cc][][1]{$a)$}
\psfrag{B}[cc][][1]{$b)$}
\psfrag{C}[cc][][1]{$c)$}

\psfrag{a}[cc][][0.7]{$\mu$}
\psfrag{b}[cc][][0.7]{$\mu,\zeta$}
\psfrag{c}[cc][][0.7]{$\widetilde{\T_B}$}
\psfrag{d}[cc][][0.7]{$\T_B$}

\begin{center}
\includegraphics[scale=0.3]{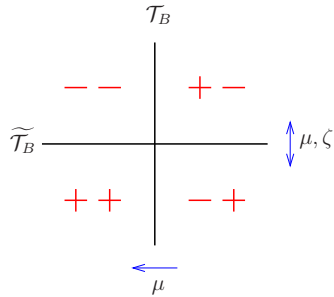}
\end{center}
\caption{Set of vacua and basis of domainwalls at the Gepner point of $X^{(8)}$.}
\label{modelsfig2}
\end{figure}

Now recalling the definitions \eqref{TA8} and \eqref{combinations} (and ignoring
RR-flux, as we said),
\begin{equation}
\eqlabel{ignoring}
\begin{split}
\T_A = \calw_{(\xi,+)} - \calw_{(\xi,-)}\,,& \qquad
\widetilde{\T_A} = \calw_{(\xi,\sigma)} - \calw_{(-\xi,\sigma)} ~,\\
\T_B = \calw_{\langle +,+\rangle} - \calw_{\langle -,+\rangle} \,,&\qquad
\widetilde{\T_B} = \calw_{\langle \mu,\zeta\rangle} - \calw_{\langle-\mu,-\zeta\rangle}~,
\end{split}
\end{equation}
we obtain an exact match of domainwall spectrum if we identify the large volume 
brane vacua $(\xi,\sigma)$ with those at the Gepner point $\langle\mu,\zeta\rangle$ 
according to
\begin{equation}
\eqlabel{finally}
\qquad\qquad \mu=\xi~,~\zeta = \xi \sigma ~.
\end{equation}
More pictorially, one may compare figure \ref{modelsfig1} one page \pageref{modelsfig1} 
with figure \ref{modelsfig2}. The former shows the collection of vacua and interpolating 
domainwalls at the large volume point in the A-model, while the second illustrates the 
domainwall spectrum obtained in the B-model at the Gepner point.

\section{Discussion and Conclusions}
\label{conclude}

In this paper, we have accumulated evidence for a mirror symmetry identification between 
A-branes defined as the real slices of one-parameter hypersurfaces in weighted projective
space and B-branes defined via certain matrix factorizations of the Landau-Ginzburg 
superpotential. We have made this identification at the level of the holomorphic data,
namely the structure of $\caln=1$ supersymmetric vacua on the D-brane worldvolume and 
the tension of BPS domainwalls between them.

The basic structure is similar to the real quintic studied in 
\cite{Walcher:2006rs,Morrison:2007bm}.  All models have in common that they possess real
Lagrangians with $H_1(L,\zet)=\zet_2$. This discrete datum corresponds to a choice of 
discrete Wilson line. Using mirror symmetry, or just based on considerations of monodromy, 
one can show that the domainwall tension separating those vacua is captured by an 
inhomogeneous Picard-Fuchs equation with inhomogeneous term $\sim z^{1/2}$. 
It is tempting to speculate that this specific type of inhomogeneous extension will generally 
describe the domainwall separating the two possible vacua of a D-brane on Lagrangians 
with $H_1(L,\Z)=\Z_2$. 

On a technical level, the key quantity to compute is the exact constant of proportionality
of the inhomogeneous term in the Picard-Fuchs equation. We have determined these constants
via two orthogonal approaches, namely consistency of monodromies and explicit computations
of Abel-Jacobi type, resulting from the B-model matrix factorizations.

The $k=8$ hypersurface differs slightly from the other models by the fact that the real
Lagrangian of interest possesses two disconnected, but homologically equivalent components,
and $H_1(L,\zet)= \zet_2\times\zet_2$. Hence, this geometry has in addition a second
discrete open string modulus corresponding to the component the D-brane is wrapped on,
as well as a second domainwall, which is formed by a D-brane on the 4-chain separating
the two components. The tension of this domainwall is simply a fractional (quantum corrected) 
closed string period. While this picture is suitable at the large volume point, we made the 
observation that continuation to the Gepner point induces a ``mixing'' of these (from a 
large volume point of view) different moduli. This is another manifestation of the break 
down of classical geometric concepts in the quantum regime, and perhaps the most
interesting lesson of our computations. We conclude with some further consequences.

\subsection{Disk Instanton Numbers}

To extract the Ooguri-Vafa invariants \cite{oova} from the Gromov-Witten expansion of the 
domainwall tension, \eqref{TA610}, we recall that the familiar $1/l^3$-multicover formula is 
replaced in the open string context by $1/l^2$. In terms of the quantum part of the domainwall 
tension \eqref{setting}, the expansion takes the form
\begin{equation}
\frac{\pi^2}{4}\;\frac{\tau(z(q))}{\varpi_0(z(q))}  = \sum_{l,d\;{\rm odd}} 
\frac{n_d^{(0,{\rm real})}}{l^2} q^{dl/2}~.
\end{equation}
The resulting integers $n_d^{(0,{\rm real})}$ (see table \ref{diskinst} for some examples) 
are BPS-invariants in the string/M-theory setup of \cite{oova}. Mathematically, they 
are predicted to be enumerative invariants counting real rational curves in $X^{(k)}$.

\begin{table}
\begin{tabular}{|c|c|}
\hline
$n_d$&disk instanton numbers for $L^{(6)}_{[M]}$\\
\hline
$1$&$24$\\
$3$&$5880$\\
$5$&$14328480$\\
$7$&$48938353176$\\
$9$&$204639347338560$\\
$11$&$965022386745454392$\\
%$13$&$4928461610491131058272$\\
%$15$&$26638444539302347587544080$\\
%$17$&$150212866666328042620525541832$\\
%$19$&$875334368623000693351386148469160$\\
\hline
\end{tabular}
\,
\begin{tabular}{|c|c|}
\hline
$n_d$&disk instanton numbers for $L^{(8)}_{[0,0,0,0,1]}$\\
\hline
$1$&$48$\\
$3$&$65616$\\
$5$&$919252560$\\
$7$&$17535541876944$\\
$9$&$410874634758297216$\\
$11$&$10854343378339853472336$\\
%$13$&$310521865321872322311676752$\\
%$15$&$9401030537961826351061423123760$\\
%$17$&$296918160618635634448828259637408528$\\
%$19$&$9690550075725044570692374881688583107408$\\
\hline
\end{tabular}

\smallskip

\begin{tabular}{|c|c|}
\hline
$n_d$&disk instanton numbers for $L^{(10)}_{[M]}$\\
\hline
$1$&$128$\\
$3$&$2886528$\\
$5$&$465626856320$\\
$7$&$112339926393132928$\\
$9$&$33254907472965538667520$\\
$11$&$11110159357336987759939410816$\\
%$13$&$4020620486167901367085148519248768$\\
%$15$&$1539912132209403478178947862140045269120$\\
%$17$&$615290814493785535045993407595549445274328448$\\
%$19$&$254042349403477846358882575420579506137171440481152$\\
\hline
\end{tabular}
\caption{Low degree BPS invariants $n^{(0,{\rm real})}_d$ for the three 
models $X^{(6)}$, $X^{(8)}$, and $X^{(10)}$.}
\label{diskinst}
\end{table}

It is interesting to note that Ooguri-Vafa integrality also holds for the second
domainwall that appears for $X^{(8)}$, see \eqref{must}. Since $\widetilde\T_A = \Pi_4/2$, where
$\Pi_4\sim \del \calf$, this integrality can be deduced from the integrality of 
ordinary closed string instanton numbers (obtained from prepotential $\calf$ with 
$1/l^3$ multi-cover formula). Note however that this is not a totally trivial check 
because of the relative factors of $2$ between open and closed string expansion.

In the absence of direct A-model computations of Gromov-Witten or Ooguri-Vafa invariants,
further checks on the enumerative predictions of table \ref{diskinst} can be derived from the 
computation of loop amplitudes in the topological string.

\subsection{One-loop test}

As explained in \cite{openbcov,unorbcov}, the domainwall tensions that we obtained as 
solutions of the inhomogeneous Picard-Fuchs equation in the previous sections constitute 
tree-level data for the computation of topological string amplitudes on the appropriate
Calabi-Yau orientifold models. Technically, we have an extension of ordinary special 
geometry to the open string sector, characterized infinitesimally by the two-point 
function on the disk, $\Delta$. This is related to the tree-level domainwall $\T$ as 
$\Delta \sim D^2 \T - C D\overline{\T}$, where $C$ is the closed string Yukawa coupling
(\ie, the infinitesimal invariant of the closed topological string), and $D$ is the
covariant derivative on moduli space. Under certain additional conditions (no contribution 
from open string moduli, tadpole cancellation, further discussed in \cite{cooy}), the 
amplitudes for higher worldsheet topology are then recursively constrained by the 
extended holomorphic anomaly equation of \cite{openbcov}, which is a generalization 
of the  BCOV equations \cite{bcov2}. The main obstacle to carrying out this program is 
the holomorphic ambiguity, which at present is not very well understood in the 
open/unoriented sector. 

For the one-loop amplitudes however, we have a complete proposal \cite{unorbcov}, 
generalizing the result of \cite{bcov1}. We can therefore just plug in the tree-level data
into this formula, and extract \cite{oova,lmv,unorbcov} one-loop BPS invariants for 
our three one-parameter hypersurfaces. One of the checks alluded to above is the following 
equality of tree-level and one-loop enumerative invariants on $X^{(6)}$:
\begin{equation}
\eqlabel{treelevelcoin}
k=6: \qquad n_1^{(0,{\rm real})} = n_2^{(1,{\rm real})} = 24~.
\end{equation} 
We view this as the real version of the coincidence of the complex enumerative invariants
(see, \eg, \cite{hkq})
\begin{equation}
\eqlabel{oneloopcoin}
k=6:\qquad n_1^{(0)} = n_2^{(1)} = 7884~,
\end{equation}
which arises from the relation between the corresponding intersection problems. The
equality \eqref{treelevelcoin} gives evidence that this relation persists in the real
version of the problem. Another check is the necessary equality of complex and real 
enumerative invariants modulo $2$, \ie,
\begin{equation}
k=6,8,10: \qquad n_d^{(\hat g,{\rm real})} = n_d^{(\hat g)} \bmod 2~,
\end{equation}
holds for all three models, all $d$, and $\hat g=0,1$.

Another interesting aspect of the loop computations derives from the disconnectedness
of the real slice of $X^{(8)}$. As observed in \cite{unorbcov}, it appears that in 
order to obtain a satisfactory BPS interpretation for open topological string amplitudes 
on compact Calabi-Yau manifold, one has to consider an orientifold model and choose a 
D-brane configuration that cancels the tadpoles. In our models, we naturally choose 
the orientifold action that we used to define the D-branes, and put exactly one D-brane 
on top of the orientifold plane. For $k=8$, however, the orientifold plane is disconnected,
and there are more tadpole cancelling D-brane configurations (ten, using just the branes
we discussed). In other words, the topological string amplitudes are a function of four 
discrete moduli $(\xi_1,\sigma_1,\xi_2,\sigma_2)$, in addition to the closed string 
modulus $t$. We have computed this function at one-loop and found an integral BPS 
expansion in all sectors. We will return to this elsewhere.

\subsection{Outlook}

The integrality of the $n_d^{(0,{\rm real})}$ from table \ref{diskinst} is a strong check 
that our overall picture is consistent. Note however that the overall normalization
of these numbers is not fixed by integrality alone (in particular, all $n_d^{(0,{\rm real})}$ 
are divisible by the first number, $n_1^{(0,{\rm real})}$). Our confidence in the enumerative
predictions therefore mainly rests on the agreement between the two different computations 
of this normalization constant, monodromy and Abel-Jacobi. As further comfort, we note that 
the corresponding predictions on the quintic \cite{Walcher:2006rs} have been verified in 
\cite{psw} using the open Gromov-Witten theory of \cite{jake} and localization on the space 
of maps to the ambient $\P^4$. It would be interesting to verify our predictions in the weighted 
case by this or other methods. 

One might also ask if similar considerations could be applied as well to Lagrangians with 
more general torsion $H_1(L,\Z)=\Z_p$. A natural guess would be that the domainwalls 
separating these vacua are similarly captured on the B-side via an inhomogeneous extension 
of the ordinary Picard-Fuchs equations of the form $\sim z^{1/p}$. It would be interesting 
to find some explicit examples which support this proposal.

\begin{acknowledgments}
We would like to thank Manfred Herbst and Wolfgang Lerche for valuable discussions. The work of D.K.\ 
is supported by an EU Marie-Curie EST fellowship. The work of J.W.\ was supported in part by 
the Swiss National Science Foundation, and by the NSF under grant number PHY-0503584.
\end{acknowledgments}

%\newpage

\appendix

\section{Inhomogeneous Picard-Fuchs equation via Griffiths-Dwork}
\label{Dwork}

In the following, we will give some more details of the main computation of section (\ref{maincompu}), i.e. the evaluation of (\ref{mainclaimeq5}). In order to be able to evaluate (\ref{mainclaimeq5}), we first need to derive the exact parts $\t\beta^{(k)}$ of the inhomogeneous Picard-Fuchs equations covering the (extended) periods of $Y^{(k)}$. We will achieve this via the Griffiths-Dwork method (see for instance \cite{Morrison:1991cd}):

The fundamental weighted homogeneous differential form of the ambient space is given by
\beq\label{Dworkeq0}
\omega^{(k)}=\sum_{i=1}^5 (-1)^{i-1}\nu_ix_i dx_1\w ...\w \widehat{dx_i} \w ...\w dx_5~,
\eq
where $\nu_i$ are the weights and $x_i$ the homogenous coordinates of the ambient weighted $\P^4$. For later convenience, we define $\omega^{(k)}_i:=\D_i\omega^{(k)}$. 

The holomorphic 3-form is given by
\beq\label{Dworkeq0a}
\Omega=Res_{W^{(k)}=0}\t\Omega~, 
\eq
with $\t\Omega=\frac{\omega^{(k)}}{W^{(k)}}$. For simplicity, the $^{(k)}$ indices are implicitly understood in the following. Then, the fundamental period
\beq
w_0=\int_\G \Omega~,
\eq
where $\G$ is usually a 3-cycle, here however we allow $\G$ to have a boundary $\D\G$,  evaluates to
\beq\label{Dworkeq0b}
w_0=\int_\G Res_{W=0}\t\Omega=\int_{T_\e(\G)}\t\Omega~,
\eq
where $T_\e$ is a small tube around $\G$.
From that we obtain
\beq
\D^l_\psi w_0=l!\int_{T_\e(\G)}\frac{(x_1x_2x_3x_4x_5)^l}{W^{l+1}}\omega~,
\eq
where we have implicitly assumed that there will be no contribution of derivatives acting on the chain. That this is indeed the case will be explicitly verified for the models under consideration. 

For $l=4$ we can express $\D_\psi^lw_0$ in terms of lower derivatives using the equations of motion $\D_i W=0$ and ``partial integration" (Griffith's reduction of pole order) and obtain in this way a differential equation of (inhomogeneous) Picard-Fuchs type satisfied by $w_0$. The calculation is lengthy, but straight-forward. 
 
\subsection{\texorpdfstring{$Y^{(6)}$}{Y(6)}}

Using the Griffiths-Dwork method as described above and the relation
\beq
\begin{split}
\p^2(1-\p^6)(x_1^4x_2^4x_3^4x_4^4x_5^4)&=\p^7(x_1^3x_2^3x_3^3x_4^3x_5^4)\D_{5}W+\p^6(x_1^2x_2^2x_3^2x_4^3x_5^5)\D_{4}W\\
&+\p^5(x_1x_2x_3^2x_4^7x_5^4)\D_{3}W+\p^4(x_2x_3^6x_4^6x_5^3)\D_{2}W\\
&+\p^3(x_2^5x_3^5x_4^5x_5^2)\D_{1}W+\p^2(x_1^4x_2^4x_3^4x_4^4x_5^2)\D_{5}W~,
\end{split}
\eq
we obtain the inhomogeneous Picard-Fuchs equation
\beq\label{Dworkeq1}
\t\L^{(6)}=\p^2(1-\p^6)\D_\p^4-2\p(1+5\p^6)\D_\p^3+(2-25\p^6)\D_\p^2-15\p^5\D_\p-\p^4=d\t\beta~,
\eq
with exact part
\beq\label{Dworkeq1a}
\begin{split}
\t\beta^{(6)}&=\frac{6}{W^4}\left[\p^7x_1^3x_2^3x_3^3x_4^3x_5^4\omega_5+\p^6x_1^2x_2^2x_3^2x_4^3x_5^5\omega_4+\p^5x_1x_2x_3^2x_4^7x_5^4\omega_3\right.\\
&+\left.\p^4x_2x_3^6x_4^6x_5^3\omega_2+\p^3x_2^5x_3^5x_4^5x_5^2\omega_1+\p^2x_1^4x_2^4x_3^4x_4^4x_5^2\omega_5\right]\\
&+\frac{2}{W^3}\left[3\p^6x_1^2x_2^2x_3^2x_4^2x_5^3\omega_5+2\p^5x_1x_2x_3x_4^7x_5^2\omega_5+3\p^6x_1^2x_2^2x_3^2x_4^3x_5^2\omega_4\right.\\
&+\p^4x_3^6x_4^6x_5\omega_5+\left.\p^5x_1x_2x_3^2x_4^7x_5\omega_3-2\p x_1^3x_2^3x_3^3x_4^3x_5\omega_5\right]\\
&+\frac{1}{W^2}\left[6\p^5x_1x_2x_3x_4^2x_5\omega_4+\p^4x_3^6x_4\omega_4+\p^5x_1x_2x_3^2x_4x_5\omega_3+x_1^2x_2^2x_3^2x_4^2\omega_5\right]\\
&+\frac{1}{W}\left[\p^4x_3\omega_3\right]~.
\end{split}
\eq

The next step in the evaluation of (\ref{mainclaimeq5}) is to define a proper tube $T_\e(\D\G)$, where we have for $Y^{(6)}$: $\D\G=C^{(6)}_+-C^{(6)}_-$ with $C^{(6)}_\zeta$ given in (\ref{curves}). For simplicity, we set $\a^{(6)}=\ii$ and drop the $^{(6)}$ indices in the following. Observe that the curves $C_\zeta$ possess two components distinguished by $\gamma=\pm 1$, i.e. $C_\zeta=C_\zeta^++C_\zeta^-$ with
\beq
C^\gamma_\zeta=\{x_1=i\zeta x_2,x_3=\ii x_4,x_5=\ii^{\frac{-\zeta-1}{2}}\gamma \sqrt{3\p}x_1x_3\}~.
\eq

Note that the curves intersect in three points: 
\beq
p^\zeta_1:=\{x_1=\ii\zeta x_2,x_3=x_4=x_5=0\},~p_2:=\{x_3=\ii x_4,x_1=x_2=x_5=0\}~.
\eq
For the same reasons as in the quintic case discussed in \cite{Morrison:2007bm}, only the neighborhood of the set of points $\{ p_1^\zeta,p^\zeta_2:=p_2\}$ gives a contribution to the integral over the tube $T_\e(\D\G)$. Hence, the integral (\ref{mainclaimeq5}) splits up into 
\beq\label{mainX6eq1}
\int_{T_\e (\D\G)}\t\beta=\sum_{\zeta,\gamma,i} \zeta\int_{T_\e(C^\gamma_\zeta;p_i^\zeta)}\t\beta~,
\eq
where we define the tubes $T_\e(C^\gamma_\zeta;p_i^\zeta)$ around $C^\gamma_\zeta$ near $p_i^\zeta$ momentarily:

The region around the points $p^\zeta_i$ is best described in the following two local charts of the ambient weighted $\P^4$:
\beq \label{mainX6eq2}
U_1:~x_i\rightarrow \frac{x_j}{x_1^{\nu_j}}~,~U_2:~x_j\rightarrow \frac{x_j}{x_3^{\nu_j}}~,
\eq
where $p^\zeta_i\subset U_i$.

Let us start with the region around $p_1^\zeta$: Changing to the inhomogeneous coordinates of $(\ref{mainX6eq2})$, i.e. 
\beq \label{mainX6eq3}
T':=\frac{x_2}{x_1}~,~X':=\frac{x_3}{x_1}~,~Y':=\frac{x_4}{x_1}~,~Z':=\frac{x_5}{x_1^2}~,
\eq
and performing subsequently the coordinate change
\beq \label{mainX6eq4}
T'\rightarrow i\zeta T~,~Y'\rightarrow -iY~,~X'\rightarrow X+Y+Z~,~Z'\rightarrow \ii^{\frac{-\zeta-1}{2}}\gamma (1+\sqrt{3\p}(Y+Z))~,
\eq
we obtain a more convenient parameterization of $C_\zeta^\gamma$ and $p^\zeta_1$:
\beq \label{mainX6eq4a}
\begin{split}
C_\zeta^\gamma&=\{T=-1,X=-Z=\frac{1}{\sqrt{3\p}}\}~,\\
p^\zeta_1&=\{T=-1,X=-Z=\frac{1}{\sqrt{3\p}},Y=0\}~.
\end{split}
\eq
Observe that the curves $C^\gamma_\zeta$ are parametrized in their respective coordinate systems by the single coordinate $Y=re^{i\phi}$, and intersects with the other curves at $r=0$. 

We now define tubes $T^{\zeta\gamma}_\e$ around the curves $C^\gamma_\zeta$ by requiring that $T^{\zeta\gamma}_\e$ lies outside of $Y^{(6)}$ inside a small neighborhood of $p_1^\zeta$ ($0\leq r<r^*$) and inside of $P^{(6)}$ else ($r\geq r^*$).
Therefore, consider the normal vectors
\beq \label{mainX6eq4b}
v_{\zeta\gamma}=a_{\zeta\gamma}D_T-\gamma \frac{\ii^{\frac{-\zeta-1}{2}}}{2}e^{-2i\phi}\D_X+\gamma \frac{\ii^{\frac{-\zeta-1}{2}}}{2}e^{-2i\phi}\D_Z~,
\eq
with
\beq\label{mainX6eq5}
a:=\frac{f(r)}{1+\gamma \ii^{\frac{\zeta+1}{2}}\sqrt{3\psi^{3}} Y^3 }~,
\eq
where $f(r)$ is a non-negative $C^\infty$ function with $f(0)=1$ and $f(r)=0$ for $r\geq r^*>0$.  Clearly $v_{\zeta\gamma}$ point inside of $P^{(6)}$ for $r>r^*$. Note that the definition of $a_{\zeta\gamma}$ naturally fixes the point $r^*$. To see that, note if we define $r^{*}=(3\psi^3)^{-1/6}$, $f(r^*)$ must vanish, since otherwise $a_{\zeta\gamma}$ would have poles at $r^{*}$. For reasons that will become clear later, we require as well that the first derivative $f'(r)$ vanishes at $r\geq r^*>0$.

One easily checks that 
\beq
D_{v_{\zeta\gamma}} W|_{C_\zeta^\gamma}=f(r)+\sqrt{3\psi^{3}}r^2>0~,
\eq
hence $v_{\zeta\gamma}$ points outside of $Y^{(6)}$ and thus we can use $v_{\zeta\gamma}$ to define proper tubes $T^{\zeta\gamma}_\e$. In detail, the tubes are parameterized in local coordinates by
\beq
T=-1+\t\e a~,~X=-Z=\frac{1}{\sqrt{3\psi}}-\gamma \frac{\ii^{\frac{-\zeta-1}{2}}}{2}\t\e e^{-2i\phi}~,
\eq
where $\t\e=e^{i\chi}\e$ and $\chi\in[0,2\pi]$.

For the evaluation of (\ref{mainX6eq1}) we need to express $\t\beta$ in the coordinates (\ref{mainX6eq4}), restrict to the respective tubes and perform the integration over 
\beq
\int_{0}^\infty dr \int_0^{2\pi}d\phi\int_0^{2\pi}d\chi~.
\eq
Since the terms occurring in $\t\beta$ are proportional to $\frac{\omega_i}{W^l}$, we especially need $\omega_i|_{T_\e^{\zeta\gamma}}$. After going to the chart $U_1$ and restricting to $X'=iY'$, we infer from (\ref{Dworkeq0}) that $\omega_i=0$ for $i\notin\{3,4\}$ and
\beq\label{mainX6eq6a}
\omega_4=-i\omega_3=dT'\w dX'\w dZ'~.
\eq
Changing to the coordinates (\ref{mainX6eq4}) then yields
\beq
dT'\w dX'\w dY'=\ii^{\frac{-\zeta+1}{2}}\zeta \gamma\sqrt{3\psi}dT\w dX\w dY~,
\eq
which restricts on the tubes $T^{\zeta\gamma}_\e$ to
\beq\label{mainX6eq7}
dT\w dX\w dY|_{T^{\zeta\gamma}_\e}=-\ii^{\frac{-\zeta-1}{2}}\gamma e^{-i\phi}a\left(1-\frac{rf'(r)}{2f(r)}\right)\t\e^2dr\w d\phi\w d\chi~.
\eq
We have now everything at our disposal to infer the contribution of the integrals over the tubes $T_\e(C^\gamma_\zeta;p_1^\zeta)$ to (\ref{mainX6eq1}). For performing the explicit calculation, note that only terms which do not come in powers of $\t\e$ can survive the integration over $d\chi$. Hence, the integration over $d\chi$ simply yields a factor of $2\pi$. For the integration over $d\phi$ it is more convenient to perform the variable transformation $e^{i\phi}\rightarrow z$, with $z\in\C$. In this coordinate, the integral becomes a line integral around the unit circle in the complex plane and only terms can contribute that have poles in the unit disk. Combined with the property $f(r)=f'(r)=0$ for $r\geq r^*$, it is easy to see that only terms can contribute which do not come in powers of $z$ in the nominator. 

Performing the explicit calculation we infer that there is no contribution from the integrals over the tubes $T_\e(C^\gamma_\zeta;p_1^\zeta)$ to (\ref{mainX6eq1}).

It remains to evaluate the contribution of $T_\e(C^\gamma_\zeta;p_2)$ to (\ref{mainX6eq1}): For that, we need to perform the same calculations as above in the coordinate chart $U_2$ which includes $p_2$. Hence, we take the inhomogeneous coordinates
\beq\label{mainX6eq8}
T':=\frac{x_4}{x_3}~,~X':=\frac{x_1}{x_3}~,~Y':=\frac{x_2}{x_3}~,~Z':=\frac{x_5}{x_3^2}~,
\eq
and perform the same coordinate redefinitions as in (\ref{mainX6eq4}) in order to obtain (\ref{mainX6eq4a}). However, this time $p^\zeta_1$ corresponds to $p_2$. Due to the symmetry of $W$, the tubes $T^{\zeta\gamma}_\e$ have the same parameterization and we can still use (\ref{mainX6eq4b})-(\ref{mainX6eq7}), if we replace 
\beq\label{mainX6eq9}
\omega_4\rightarrow \omega_2~, ~\omega_3\rightarrow \omega_1~.
\eq

Performing the explicit calculation similar as in chart $U_1$, we infer that we obtain contributions from the term
\beq
\int_{{T_\e(C^\gamma_\zeta;p_2)}}\frac{6 \psi^4 x_2 x_3^6 x_4^6 x_5^3 \omega_2}{W^4}=\zeta\frac{2}{3}\pi^2.
\eq
Note that we have taken an additional normalization factor of $6^{-1}$ due to $G_{\rm fix}$ given in table \ref{invgps} into account ($p_2$ is a singular point).

It remains to show that the underlying assumption that we have no contribution from derivatives acting on $T_\e(\G)$ indeed holds. The argumentation is as in \cite{Morrison:2007bm}. For that, note that the normal vectors implementing first order deformations of $C_\zeta^\gamma$ are given by
\beq
n_{\zeta\gamma}=-\ii^{\frac{\zeta+1}{2}}\gamma \frac{x_5}{\sqrt{3\p}}\frac{1}{2\p}(\D_3-\ii\D_4)~. 
\eq
Hence, we have that
\beq
n_{\zeta\gamma} \omega=-\ii^{\frac{\zeta+1}{2}}\gamma \frac{x_5}{\sqrt{3\p}}\frac{1}{2\p}(\omega_3-\ii\omega_4)=0~,
\eq
where we used (\ref{mainX6eq6a}).

\subsection{\texorpdfstring{$Y^{(8)}$}{Y(8)}}

The discussion of the remaining two models $Y^{(8)}$ and $Y^{(10)}$ is very similar to $Y^{(6)}$, hence we will be brief:

For $Y^{(8)}$ we use the relation
\beq
\begin{split}
\p^3(1-\p^8)(x_1^4x_2^4x_3^4x_4^4x_5^4)&=\p^{10}(x_1^3x_2^3x_3^3x_4^3x_5^4)\D_{5}W+\p^9(x_1^2x_2^2x_3^2x_4^3x_5^4)\D_{4}W\\
&+\p^8(x_1x_2x_3^2x_4^9x_5^3)\D_{3}W+\p^7(x_2x_3^8x_4^8x_5^2)\D_{2}W\\
&+\p^6(x_2^7x_3^7x_4^7x_5)\D_{1}W+\p^5(x_1^6x_2^6x_3^6x_4^6x_5)\D_{5}W\\
&+\p^4(x_1^5x_2^5x_3^5x_4^5x^2_5)\D_{5}W+\p^3(x_1^4x_2^4x_3^4x_4^4x_5^3)\D_{5}W~,
\end{split}
\eq
to obtain the inhomogeneous Picard-Fuchs equation 
\beq\label{Dworkeq2}
\t\L^{(8)}=\p^3(1-\p^8)\D^4_\p-\p^2(6+10\p^8)\D^3_\p+5\p(3-5\p^8)\D^2_\p-15(1+\p^8)\D_\p-\p^7=d\t\beta~,
\eq
with exact part
\beq
\begin{split}\label{Dworkeq2a}
\t\beta^{(8)}&=\frac{6}{W^4}\left[\p^{10}x_1^3x_2^3x_3^3x_4^3x_5^4\omega_5+\p^9x_1^2x_2^2x_3^2x_4^3x_5^4\omega_4+\p^8x_1x_2x_3^2x_4^9x_5^3\omega_3+\p^7x_2x_3^8x_4^8x_5^2\omega_2\right.\\
&+\left.\p^6x_2^7x_3^7x_4^7x_5\omega_1+\p^5x_1^6x_2^6x_3^6x_4^6x_5\omega_5+\p^4x_1^5x_2^5x_3^5x_4^5x_5^2\omega_5+\p^3x_1^4x_2^4x_3^4x_4^4x_5^3\omega_5\right]\\
&+\frac{2}{W^3}\left[3\p^9x_1^2x_2^2x_3^2x_4^2x_5^3\omega_5+2\p^8x_1x_2x_3x_4^9x_5^2\omega_5+2\p^9x_1^2x_2^2x_3^2x_4^{10}x_5\omega_5\right.\\
&\left.+2\p^{10}x_1^3x_2^3x_3^3x_4^4x_5\omega_4-2\p^{10}x_1^3x_2^3x_3^3x_4^3x_5^2\omega_5+\p^7x_3^8x_4^8x_5\omega_5+\p^8x_1x_2x_3^9x_4^2x_5\omega_4\right.\\
&\left.+\p^{9}x_1^2x_2^2x_3^3x_4^2x_5^2\omega_3-6\p^{2}x_1^3x_2^3x_3^3x_4^3x_5^2\omega_5-3\p^{3}x_1^4x_2^4x_3^4x_4^4x_5\omega_5-\p^{4}x_1^5x_2^5x_3^5x_4^5\omega_5\right]\\
&+\frac{1}{W^2}\left[4\p^8x_1x_2x_3x_4^2x_5\omega_4+2\p^9x_1^2x_2^2x_3^2x_4^3\omega_4-6\p^9x_1^2x_2^2x_3^2x_4^2x_5\omega_5+\p^7x_3^8x_4\omega_4\right.\\
&+\left.3\p^8x_1x_2x_3^2x_4x_5\omega_3+15\p x_1^2x_2^2x_3^2x_4^2x_5\omega_5+3\p^2x_1^3x_2^3x_3^3x_4^3\omega_5 \right]\\
&+\frac{1}{W}\left[\p^7x_3\omega_3-15x_1x_2x_3x_4\omega_5\right]~.
\end{split}
\eq

In order to evaluate (\ref{mainclaimeq5}), we need to define proper tubes around the curves $C^{(8)}_{\langle \mu,\zeta\rangle}$. For simplicity, in the following we will just discuss the $C^{(8)}_{\langle -,\zeta\rangle}$ part. The $C^{(8)}_{\langle +,\zeta\rangle}$ part can be discussed similarly and the results just differ by an overall sign\footnote{\label{generalprinciple} In fact, this must be so since by general principles, the total intersection of the hypersurface with a plane should give a vanishing result.} (therefore the $\mu$ in (\ref{intresults})). Hence, let us consider the curves
\beq
C_{\langle-, \zeta\rangle}:=\{x_1=\a^\zeta x_2,x_3=\a x_4,x_5= 2\a^{-\zeta-1}\p x_1^2x_3^2\}~.
\eq
In the following we will denote these curves simply as $C_{\zeta}$ and we will implicitly set $\a=e^{i\pi/8}$. The two curves intersect in the point
\beq
p:=\{x_3=\a x_4,x_1=x_2=x_5=0\}~.
\eq
As a consequence, (\ref{mainX6eq1}) reduces for $Y^{(8)}$ to 
\beq\label{mainX8maineq}
\int_{T_\e (\D\G)}\t\beta=\sum_{\zeta} \zeta\int_{T_\e(C_\zeta;p)}\t\beta~.
\eq

For the evaluation of (\ref{mainX8maineq}), we go to the local chart $U_2$ defined in (\ref{mainX6eq8}) and perform the variable redefinitions
\beq\label{mainX8eq3}
T'\rightarrow -\alpha^{-1}T~,~Y'\rightarrow \alpha^{-\zeta}Y~,~X'\rightarrow X+Y+Z~,~Z'\rightarrow \alpha^{-\zeta-1}(1+\sqrt{2\p}(Y+Z))^2~.
\eq
Then, we have
\beq
\begin{split}
C_\zeta&=\{T=-1,X=-Z=\frac{1}{\sqrt{2\p}}\}~,\\
p&=\{T=-1,X=-Z=\frac{1}{\sqrt{2\p}},Y=0\}~.
\end{split}
\eq
The tubes $T^\zeta_\e$ around $C_\zeta$ are parameterized similar as for $Y^{(6)}$ via normal vectors $v_\zeta$ defined by
\beq\label{mainX8eq4}
v_\zeta=a_\zeta\D_T-\frac{\ii}{4}\a^{2(\zeta-1)}e^{-3\ii\phi}\D_X+\frac{\ii}{4}\a^{2(\zeta-1)}e^{-3\ii\phi}\D_Z~,
\eq
with
\beq
a_\zeta:=\frac{f(r)}{1-2\ii\zeta\a^{2(\zeta-1)}\p^2 Y^4 }~.
\eq
This choice ensures that 
\beq
D_{v_{\zeta}}W|_{C_\zeta}=f(r)+\p^3r^3>0~,
\eq
such that $T^\zeta_\e$ is well defined.

Thus, the tubes $T^\zeta_\e$ are locally parameterized by
\beq
T=-1+\t\e a~,~X=-Z=\frac{1}{\sqrt{2\p}}-\t\e\frac{\ii}{4}\a^{2(\zeta-1)}e^{-3\ii\phi}~,
\eq
where $\t\e=e^{\ii\chi}\e$.

As a last piece, we need the restriction of the forms $\omega_i$ to $T^\zeta_\e$: Going to the chart $U_2$ and restricting to $X'=\a^\zeta Y'$, we directly infer that $\omega_i=0$ for $i\notin\{1,2\}$ and that
\beq\label{mainX8eq7a}
\omega_2=-\a^\zeta \omega_1=dT'\w dX'\w dZ'~.
\eq
Changing to the coordinates (\ref{mainX8eq3}) then yields
\beq
dT'\w dX'\w dZ'=-2\a^{-2-\zeta}\sqrt{2\psi}(1+ \sqrt{2\psi}(Y+Z))dT\w dX\w dY~.
\eq
Further,
\beq
dT\w dX\w dY|_{T^\zeta_\e}= -\ii\frac{3}{4}\a^{2(\zeta-1)}a e^{-2\ii\phi}\t\e^2\left(1-\frac{rf'(r)}{3f(r)}\right)dr\w d\phi\w d\chi~.
\eq
Thus,
\begin{multline}
\label{mainX8eq8}
\omega_2|_{T^\zeta_\e}=-\a^\zeta\omega_1|_{T^\zeta_\e}=\\ =\ii3\a^{-4+\zeta}a \psi e^{-2\ii\phi}\left(re^{\ii\phi}+ \frac{\ii}{4}\a^{2(\zeta-1)}e^{-3\ii\phi}\t\e\right)\t\e^2\left(1-\frac{rf'(r)}{3f(r)}\right)dr\w d\phi\w d\chi~.
\end{multline}

We have now everything at our disposal to calculate (\ref{mainX8maineq}). After performing the calculations, we infer that we have a contribution from the term
\beq
\int_{T_\e(C_\zeta;p)}\frac{6\psi^6 x_2^7 x_3^7 x_4^7 x_5 \omega_1}{W^4}=\zeta 3\pi^2\psi^2~,
\eq
where we included an additional normalization factor of $4^{-1}$, similar as in the $Y^{(6)}$ case.

Similar as for X(6), we infer from the normal vector
\beq
n_\zeta=-\frac{\a^{\zeta+1}}{4\psi^2}\frac{x_5}{x_1}(\D_1+\a^{-\zeta}\D_2)~. 
\eq
that 
\beq
n_\zeta\omega=-\frac{\a^{\zeta+1}}{4\psi^2}\frac{x_5}{x_1}(\omega_1+\a^{-\zeta}\omega_2)=0~,
\eq
where we used (\ref{mainX8eq7a}). Hence, the underlying assumption that we have no contribution from derivatives acting on $T_\e(\G)$ holds. 

\subsection{\texorpdfstring{$Y^{(10)}$}{Y(10)}}

For $Y^{(10)}$ we use the relation
\beq
\begin{split}
\p^3(1-\p^{10})(x_1^4x_2^4x_3^4x_4^4x_5^4)&=\p^{12}(x_1^3x_2^3x_3^3x_4^3x_5^4)\D_{5}W+\p^{11}(x_1^2x_2^2x_3^2x_4^3x_5^4)\D_{4}W\\
&+\p^{10}(x_1x_2x_3^2x_4^6x_5^3)\D_{3}W+\p^9(x_2x_3^{10}x_4^5x_5^2)\D_{2}W\\
&+\p^8(x_2^9x_3^9x_4^4x_5)\D_{1}W+\p^7(x_1^8x_2^8x_3^8x_4^3x_5)\D_{5}W\\
&+\p^6(x_1^7x_2^7x_3^7x_4^2x^2_5)\D_{5}W+\p^5(x_1^6x_2^6x_3^6x_4x_5^3)\D_{5}W\\
&+\p^4(x_1^5x_2^5x_3^5x^4_5)\D_{5}W+\p^3(x_1^4x_2^4x_3^4x_5^4)\D_{4}W~,
\end{split}
\eq
to obtain the inhomogeneous Picard-Fuchs equation 
\beq\label{Dworkeq3}
\t\L^{(10)}=\p^3(1-\p^{10})\D^4_\p-10\p^2(1+\p^{10})\D^3_\p+5\p(7-5\p^{10})\D^2_\p-5(7+3\p^{10})\D_\p-\p^9=d\t\beta~,
\eq
with exact part
\beq
\begin{split}\label{Dworkeq3a}
\t\beta^{(10)}&=\frac{6}{W^4}\left[\p^{12}x_1^3x_2^3x_3^3x_4^3x_5^4\omega_5+\p^{11}x_1^2x_2^2x_3^2x_4^3x_5^4\omega_4+\p^{10}x_1x_2x_3^2x_4^6x_5^3\omega_3\right.\\
&+\left.\p^9x_2x_3^{10}x_4^5x_5^2\omega_2+\p^8x_2^9x_3^9x_4^4x_5\omega_1+\p^7x_1^8x_2^8x_3^8x_4^3x_5\omega_5+\p^6x_1^7x_2^7x_3^7x_4^2x_5^2\omega_5\right.\\
&+\left.\p^5x_1^6x_2^6x_3^6x_4x_5^3\omega_5+\p^4x_1^5x_2^5x_3^5x_5^4\omega_5+\p^3x_1^4x_2^4x_3^4x_5^4\omega_4\right]\\
&+\frac{2}{W^3}\left[3\p^{11}x_1^2x_2^2x_3^2x_4^2x_5^3\omega_5+2\p^{10}x_1x_2x_3x_4^6x_5^2\omega_5+2\p^{11}x_1^2x_2^2x_3^2x_4^{3}x_5^2\omega_4\right.\\
&\left.+\p^{9}x_3^{10}x_4^5x_5\omega_5 + \p^{10}x_1x_2x_3^{11}x_4^2x_5\omega_4+\p^{11}x_1^2x_2^2x_3^3x_4^2x_5^2\omega_3+7\p^5x_1^6x_2^6x_3^6x_4x_5\omega_5\right.\\
&\left.-10\p^{2}x_1^3x_2^3x_3^3x_4^3x_5^2\omega_5-10\p^{3}x_1^4x_2^4x_3^4x_4^4x_5\omega_5-10\p^{4}x_1^5x_2^5x_3^5x_4^5\omega_5\right.\\
&\left.-10\p^{5}x_1^6x_2^6x_3^6x_4^2\omega_4-\p^{6}x_1^7x_2^7x_3^7x_4^2\omega_5\right]\\
&+\frac{1}{W^2}\left[5\p^{10}x_1x_2x_3x_4^2x_5\omega_4+2\p^{10}x_1x_2x_3^2x_4x_5\omega_3+35\p x_1^2x_2^2x_3^2x_4^2x_5\omega_5+\p^9x_3x_4^5\omega_3\right.\\
&+\left.5\p^2x_1^3x_2^3x_3^3\omega_5+5\p^3 x_1^4x_2^4x_3^4\omega_4+13\p^4x_1^5x_2^5x_3^5\omega_5+10\p^2x_1^3x_2^3x_3^3x_4^3\omega_5 \right]\\
&+\frac{1}{W}\left[\p^9x_4\omega_4-35x_1x_2x_3x_4\omega_5\right]~.
\end{split}
\eq

It is easy to see that the curve for $Y^{(10)}$ given in (\ref{curves}) can be obtained by the intersection $P'\cap Y^{(10)}$, with the plane
\beq
P':=\{x_1=\alpha^\zeta x_2, \frac{1}{\sqrt{5}}\alpha^5 x_3^5+x_5-\psi x_1x_2x_3x_4=0 \}.
\eq
By the principle mentioned in the footnote on page \pageref{generalprinciple}, we can (up to a sign) use either of the components of $P'\cap Y^{(10)}$, \ie, either \eqref{curves} or the curve $x_4^2=0$. The latter curve is however also contained in the intersection of $Y^{(10)}$ with some different planes, and for incidental reasons, we prefer to present the evaluation of the integral in one of these alternative planes. More explicitly, the curve from \eqref{curves} is equivalent to the curve
\beq
C^\gamma_\zeta:=\{x_1=i\zeta x_2,x_5=i\frac{1}{5^{1/2}}x_3^5, x_4^2=\gamma \sqrt{\zeta}\sqrt{5^{1/2}\p}x_1x_3^3\}~.
\eq
with $\gamma=\pm 1$, which we obtain from $P\cap Y^{(10)}$, where $P$ is related to $P'$ via the coordinate transformation $x_5\rightarrow x_5+\psi x_1x_2x_3x_4$. Similar as the curve for $Y^{(6)}$, the $C_\zeta$ curve splits into two components, i.e. $C_\zeta=C^+_\zeta+C^-_\zeta$.

The component curves meet in three points:
\beq
p_1^\zeta=\{x_1=i\zeta x_2,x_3=x_4=x_5=0\},~p_2=\{x_5=i\frac{1}{5^{1/2}}x_3^5,x_1=x_2=x_4=0\}~.
\eq

In order to evaluate (\ref{mainX6eq1}), we go to the local chart $U_1$ given by (\ref{mainX6eq3}) and  perform the following variable redefinitions:
\begin{multline}
\label{mainX10eq1}
T'\rightarrow i\zeta T~,~Y'\rightarrow i5^{-1/2}Y^{10}~,~X'\rightarrow (X+Y+Z)^2~,\\
~Z'\rightarrow \gamma^{1/2}\zeta^{1/4}(1+5^{1/24}\p^{1/12}(Y+Z))^{3}~.
\end{multline}
Then,
\beq
\begin{split}
C_{\zeta}^\gamma&=\{T=-1,X=-Z=\frac{1}{\sqrt{5^{1/12}\p^{1/6}}}\}~,\\
p_1^\zeta&=\{T=-1,X=-Z=\frac{1}{\sqrt{5^{1/12}\p^{1/6}}},Y=0\}~.
\end{split}
\eq

We define the tubes $T^{\zeta\gamma}_\e$ via the normal vectors
\beq
v_{\zeta\gamma}=a_{\zeta\gamma}\D_T-\frac{\gamma^{3/2}\zeta^{5/4}}{12\cdot 5^{-3/8}}e^{-14i\phi}\D_X+\frac{\gamma^{3/2}\zeta^{5/4}}{12\cdot 5^{-3/8}}e^{-14i\phi}\D_Z~,
\eq
with
\beq
a_{\zeta\gamma}:=\frac{f(r)}{1+\gamma^{1/2}\zeta^{3/4} 5^{-3/8}\p^{5/4} Y^{15} }~.
\eq
Then, we have
\beq
D_{v_{\zeta\gamma}}W|_{C_\zeta^\gamma}=f(r)+\p^{5/4}r^{14}>0~,
\eq
such that the tubes $T^{\zeta\gamma}_\e$ are well defined and locally parameterized by
\beq
T=-1+\t\e a~,~X=-Z=\frac{1}{\sqrt{5^{1/12}\p^{1/6}}}-\frac{\gamma^{3/2}\zeta^{5/4}}{12\cdot 5^{-3/8}}e^{-14i\phi}\t\e~,
\eq
where $\t\e=e^{i\chi}\e$.

Let us now consider the forms $\omega_i$: Going to the chart $U_1$ and restricting to $Y'= \ii 5^{-1/2}X'^5$, we directly infer that $\omega_i=0$ for $i\notin\{3,5\}$ and that
\beq
\omega_3=-dT'\w dY'\w dZ'~,~\omega_5=-dT'\w dX'\w dZ'~.
\eq
Thus,
\beq\label{mainX10eq8a}
w_5=-\ii5^{-1/2}X'^{-4}w_3=-dT'\w dX'\w dZ'~.
\eq
Changing to the coordinates (\ref{mainX10eq1}) then yields
\beq
dT'\w dX'\w dZ'=\ii \gamma^{1/2}\zeta^{1/4} 6\cdot  5^{1/24}\p^{1/12}(X+Y+Z)(1+5^{1/24}\p^{1/12}(Y+Z))^2dT\w dX\w dZ~.
\eq
Further,
\beq
dT\w dX\w dZ|_{T_\e^{\zeta\gamma}}=- \frac{7}{6} 5^{3/8}\gamma^{3/2}\zeta^{5/4}a e^{-13i\phi}\t\e^2\left(1-\frac{rf'(r)}{14f(r)}\right)dr\w d\phi\w d\chi~.
\eq
Hence,
\beq
\begin{split}
w_5|_{T^{\zeta\gamma}_\e}&=i7\cdot5^{3/24}\p^{1/12}a e^{-13i\phi}Y(1+5^{1/24}\p^{1/12}(Y+Z))^2\!\left(\!1\!-\frac{rf'(r)}{14f(r)}\right)\t\e^2dr\w d\phi\w d\chi,\\
w_3|_{T^{\zeta\gamma}_\e}&=7\cdot5^{15/24}\p^{1/12}a e^{-13i\phi}Y^9(1+5^{1/24}\p^{1/12}(Y+Z))^2\!\left(\!1\!-\frac{rf'(r)}{14f(r)}\right)\t\e^2dr\w d\phi\w d\chi.
\end{split} 
\eq
After performing the explicit integration, we infer that we obtain contributions from the following terms occurring in $\t\beta^{(10)}$:
\beq\label{mainX10eq4}
\begin{split}
\int_{T^{\zeta\gamma}_\e}\frac{6 \psi^{6} x_1^7 x_2^7 x_3^7 x_4^2 x_5^2 \omega_5}{W^4}&= \zeta\frac{2}{5} \sqrt{5} \pi^2 \psi^3~,\\
\int_{T^{\zeta\gamma}_\e}\frac{14 \psi^5  x_1^6 x_2^6 x_3^6 x_4 x_5  \omega_5 }{W^3}&=- \zeta\frac{7}{5} \sqrt{5} \pi^2 \psi^3~,\\
\int_{T^{\zeta\gamma}_\e}\frac{13 \psi^4  x_1^5 x_2^5 x_3^5     \omega_5} {W^2}&=\zeta\frac{13}{5}  \sqrt{5} \pi^2 \psi^3~,
\end{split}
\eq
where we have included an additional normalization factor of $10^{-1}$.

With similar computations as above one can show that for chart $U_2$ (which includes $p_2$) no contribution arises.

Summing the contributions given in (\ref{mainX10eq4}), we infer
\beq
\int_{T_\e(C_\zeta^\gamma ; p^\zeta_1)}\t\beta=\zeta\frac{8}{5} \sqrt{5} \pi^2 \psi^3~.
\eq

It remains to show that the underlying assumption is correct: We have
\beq
n_{\zeta\gamma}=\gamma^{-1}\zeta^{-1/2} \frac{1}{4}5^{-1/4}\psi^{-3/2}\frac{x_4^2}{x_3}\left(\D_3-i5^{1/2}x_3^4\D_5\right)~.
\eq
Hence,
\beq
n_{\zeta\gamma}\omega=\gamma^{-1}\zeta^{-1/2} \frac{1}{4}5^{-1/4}\psi^{-3/2}\frac{x_4^2}{x_3}\left(\omega_3-i5^{1/2}x_3^4\omega_5\right)=0~,
\eq
where we used (\ref{mainX10eq8a}) .

\end{document}